\documentclass[prc,12p,showpacs,showkeys]{revtex4-1}
\usepackage{graphicx}
\usepackage{dcolumn}
\usepackage{longtable}
\usepackage{amsmath}
\usepackage{amssymb}
\usepackage{natbib}
\usepackage{array}
\usepackage{multirow}
\usepackage{calligra}
\usepackage{float}
\usepackage{commath}

\begin{document}
	\title{Conformal symmetry in nuclear structure}
	\author{Panagioti E Georgoudis}
	\affiliation{Grand Accelerateur National d'Ions Lourds, CEA/DRF-CNRS/IN2P3, BP 55027, F-14076 Caen Cedex 5, France}
	\email{panagiotis.georgoudis@ganil.fr}
	\date{June 30, 2021}
\begin{abstract}
The purpose of this paper is to introduce the unitary limit as applied in systems of cold atoms into collective states of heavy, even-even nuclei and to identify a related physical example. This is accompanied by the determination of observables of conformal symmetry in nuclear structure. A Hamiltonian is defined that governs the scattering process of an incident pair of two slow neutrons onto a heavy, even-even target nucleus in the framework of the Interacting Boson Model of nuclear structure. A unitary pair-collective state interaction is introduced along with the effective range expansion of a pair-collective state scattering length. The solutions to the coupled channels equations provide a scattering state for the two neutrons that couples with intermediate states of the $A+2n$ compound nucleus. The unitary limit manifests itself when that intermediate states correspond to pair-collective state resonances the positions of which coincide with the energies of IBM states of the closed channels. The position and the width of each resonance is measurable via the fluctuation of the cross section which tunes the pair-collective state scattering length. Conformal symmetry is represented in a tower of equally spaced states that emerges at the unitary limit from two-boson excitations on the IBM state of the closed channel. These tower states are experimentally testable as a regularity pattern of fluctuations of the cross section, in terms of regular positions and widths. The manifestation of the unitary limit and conformal symmetry via fluctuations of the cross section indicates the $A+2n$ compound nucleus as a physical laboratory for the examination of the BCS-BEC crossover, of an underlying critical point and of algebras with infinite number of generators.
\end{abstract}
	
	\maketitle
\section{Introduction}\label{s1}
A challenge in the current theoretical description of low-lying collective states of atomic nuclei is the explicit manifestation of the strong interactions at the level of mesons or of Quantum Chromo Dynamics. However, atomic nuclei exhibit symmetries in their collective states that are hosted in the group theoretical framework of the Interacting Boson Model \cite{IBM} with remarkable phenomenological success. At present, large scale shell model calculations extract nuclear collective states in the framework of effective field theories after the projection on a symmetry basis. In general, one may raise the question about the relation of the symmetries of low-lying collective states of atomic nuclei with the symmetries of the strong interactions, either with the SU(N) gauge group or with conformal symmetry as the classical limit of QCD. This problem reflects in part the more general one of the understanding of the relation between the symmetries manifested in stationary states of subatomic structures, of which atomic nuclei are an example, with the symmetries of the fundamental interactions as they are manifested in scattering states between the constituents of those structures. 

Notwithstanding the complexity of the nuclear many body problem, spectra and E2 transition rates of low-lying collective states of heavy nuclei with even number of neutrons and even number of protons (even-even) exhibit remarkable regularities that are classified in terms of three dynamical symmetry limits of the IBM. Three different subgroup chains of the $U(6)$ group of the IBM consist of the dynamical symmetry limits and correspond to shapes of the nuclear surface. These are the $U(5)$ limit (spherical shape), the $O(6)$ limit  ($\gamma$ unstable shape) and the $SU(3)$ limit (axially symmetric shape). However, there are transition regions between these shapes that manifest critical points of Quantum Phase (symmetry of the ground state) Transitions between the dynamical symmetry limits of the IBM. For example, some even-even nuclei manifest a shape between spherical and $\gamma$ unstable that is classified according to the $E(5)$ critical point symmetry \cite{E(5)}. In nuclear structure and particularly in heavy even-even nuclei, the occurrence of conformal invariance is implied through the $E(5)$ symmetry that corresponds to the critical point of a 2nd order Quantum Phase Transition.

 The $U(6)$ group of the IBM is the symmetry of a six dimensional harmonic oscillator. A spin zero $s$ boson and a spin two $\bf{d}$ boson are its building blocks. The $s$ and $\bf{d}$ bosons are identified as valence nucleon pairs of total angular momentum zero and two respectively. An overall fixed boson number $N_{b}$ determines an atomic nucleus.   

In non-relativistic quantum mechanics, unitary fermions manifest non-relativistic conformal symmetry and support a quantum critical point \cite{Randeria, Son}. This is the unitary limit - an otherwise scattering problem at the limit of infinite scattering length where the scattering amplitude becomes maximum that supports a bound state of zero energy. The unitary limit is experimentally accessible in systems of cold atoms via Feshbach resonances \cite{WC}. The same limit has been theoretically explored in light nuclei \cite{Konig} and since it reflects universality one may expect that its application in nuclear structure should not distinguish between light and heavy nuclei. Feshbach resonances in systems of cold atoms, the physical situation that hosts the unitary limit, exemplify the occurence of intermediate states of the Feshbach formalism \cite{Timmermans} which however were initially introduced in compound nuclei \cite{Feshbach}. 

The primary purpose of this paper is the connection of low-lying collective states of heavy and even-even nuclei with the unitary limit and the identification of a related physical example. Even though such a purpose may be theoretically explored via methods used in light nuclei, e.g via the No Core Shell Model \cite{Konig}, the context of a boson model for nuclear structure i.e the IBM is chosen here. The rationale for this choice is that the IBM is amenable to a symmetry-based approach to the unitary limit that has been applied in Bose-Einstein condensates of cold atoms at the vicinity of a Feshbach resonance \cite{WC}. In the physics of cold atoms, this approach makes use of the $SO(2,1)$ group which classifies collective states of cold atoms in a time-dependent trap and is isomorphic to the conformal group in one dimension - time.
 In nuclear physics, a realization of this approach is presented here which combines the Feshbach formalism with the IBM. The related physical example refers to the time-dependent process of the formation (decay) of a heavy compound, even-even nucleus. As a result, the applicability of the IBM extends to compound nuclei and the representations of the one dimensional conformal group are introduced in nuclear structure. The connection to the physical experiment is identified on the fluctuations of cross sections of compound nuclei.

The formation (decay) of the compound nucleus under examination results from the scattering process of an incident (outgoing) pair of neutrons onto (from) a heavy and even-even target nucleus. Such a process is the outcome of the algebraic comparison between the Schrodinger equation used in trapped cold atoms with the Schrodinger equation of the $O(6)$ limit of the IBM. That algebraic comparison provides a symmetry-based framework for the construction of the states of the reaction channels and of the Hamiltonian. Each reaction channel is composed of a target state written in the IBM $O(6)$ basis multiplied with the scattering state of the pair. The terms of the Hamiltonian are determined for that particular situation which leads to a resonance. The $O(6)$ symmetry and a very low energy of the incident (outgoing) pair leads to the approximation of the interaction between the incident neutrons and the collective state as a pair-collective state coupling. The resulted resonance is formed (decays) by the coupling of the incident (outgoing) $s$-wave neutron pair with the collective ground state of the target.

For the unitary limit, the  pair-collective state coupling as well as interactions between the incident neutrons themselves are determined in analogy with unitary interactions that govern  the scattering of cold atoms in a dilute atomic gas. The application of the IBM to the compound nucleus leads to the consideration of the collective nuclear state as a gas of nucleons that form correlated pairs at low temperatures. In nuclear physics, the notion of cold applies to low-lying resonances in compound nuclei. Their formation (decay) occurs when the energy of the incident (outgoing) two neutrons is very low i.e very close to the two-neutron separation energy and therefore the nuclear temperature of the compound nucleus is interpreted as cold \cite{Blatt}. The notion of dilute stands for the valence space of the heavy, even-even nucleus where the IBM emerges and implies a range for the inter-nucleon separations much larger than the range of the strong force. The pair-collective state coupling is then approximated by a delta interaction. This interaction is the analog of the unitary molecule-molecule coupling in systems of cold fermionic atoms that form diatomic molecules of bosonic character at low temperatures \cite{Pricoupenko}.

Resonances in compound nuclei constitute the so-called intermediate states which are detectable via fluctuations of the cross section \cite{Feshbach, Weidenmuller}. There is a cross section for the elastic reaction to which a fluctuation adds due to the compound system. This is the case of a compound-elastic reaction realized when the exit channel of the compound system is the same as the entrance one. The fluctuations - averaged over an energy interval - manifest the widths and the positions of the resonances that in  general are random \cite{Weidenmuller}. However, the IBM and the representations of the one-dimensional conformal group give determined positions and widths for the resonances of the pair-collective state system.  It turns out that such a determined fluctuation tunes the pair-collective state scattering length and the latter becomes infinite at resonance manifesting the unitary limit. Conformal symmetry produces a regularity pattern of a sequence of fluctuations of the cross section in terms of regular positions and widths of the resonances in contrast with their usual random appearance.

In general, the representations of the $SO(2,1)$ conformal group should not be restricted to only one type of a scattering or reaction process involving two slow neutrons as the projectiles or reaction products from a heavy target. Decays with two neutrons occur in the damping of giant resonances in photoneutron ($\gamma,2n$) reactions as for instance in a giant dipole resonance that occurs in ultraperipheral reactions of heavy ions and in ($\alpha,2n$) reactions in the damping of giant monopole resonances which manifest compressibility effects \cite{Berman}. In each case of the particular experiment used, the Hamiltonian of the compound system needs to be specified and this affects also the pair-collective state coupling which is responsible for the positions and the widths of the pair-collective state resonances.

As a tentative application of the representations of the one dimensional conformal group in the IBM-compound Hamiltonian one examines the case of a two-neutron transfer reaction. This is compatible with the simplest form of the $O(6)$ IBM Hamiltonian used here and serves as a candidate for the low energy region of the unitary limit. In a two-neutron transfer, the pair-collective state coupling needs to incorporate a second extra term apart from the unitary interaction. This extra term represents the impact of the ingoing neutron pair - as the external field - on the IBM target states and affects their boson number. It is expressed as the creation of one boson to form the resonance as well as the annihilation of one boson for the subesquent decay of the resonance. This term is the two-neutron transfer operator of the IBM applied however to the situation of a compound nucleus as the superposition of the creation of one boson and its subsequent annihilation. In analogy with Feshbach resonances in systems of cold atoms, that extra term reflects the external "magnetic" or gauge field which couples two different reaction channels. 

A physical process is therefore determined for the examination of the unitary limit in heavy nuclei. This is a two-neutron transfer which however creates a short lived compound $A+2n$ system that subsequently decays by ejecting the neutron pair back. This type of reaction may be realized in short-lived and neutron rich exotic isotopes which are expected to emit back the transferred neutron pair in a relatively short time due to their unstable character. In contrast to a direct reaction, here the amplitude of the two-neutron transfer appears as a fluctuation to the cross section in a process typical for resonances in compound nuclei. The width as well as the position of the resonance are provided for comparison with the experiment. 

However, the $SO(2,1)$ conformal group is represented in the states of the IBM-compound Hamiltonian through a tower of equally spaced states. The ground member of the tower is the intermediate state from the creation of one boson on the target state. The excited members of the tower are two-boson states on the top of the intermediate state. They  are manifested as a regularity pattern of the fluctuations of the compound-elastic cross section, regular positions and widths of the fluctuations with respect to variations of the mass number of the target nucleus or with respect to the number of incident neutron pairs. These members resemble pre-equilibrium contributions to the compound nucleus \cite{Feshbach2} which here apply as pre-equilibrium contributions to the two-neutron transfer. 

Regularities of stationary collective states of heavy and even-even nuclei are reproduced with remarkable success by the first version of the IBM (IBM-1) in which no distinction with respect to the isospin occurs for the valence nucleons. There are other versions of the IBM such as the IBM-2 \cite{IBM} where the valence nucleons are distinguished with respect to isospin. However, the functionality of the IBM-1 with respect to even-even nuclei is traditionally interpreted as the emergent simplicity out of complexity \cite{Casten} of the nuclear many-body problem as that simplicity is reflected on the symmetries of the low-lying collective states. Compound nuclei present an even more challenging case of complexity. The application of the IBM-1 combined with the Feshbach formalism in such a case, reveals again regularity patterns that are associated with conformal symmetry. These patterns may be tested experimentally through the fluctuations of averages of cross sections in compound nuclei.

This section continues with a brief reminder of the unitary limit.

\subsection{Unitary limit}\label{s2}
Take a system of N particles of arbitrary spin with 3N coordinates $r_{i}$, $r_{ij}=r_{i}-r_{j}$, each of mass $m$ subjected to the Hamiltonian
\begin{equation}\label{e1}
	H=\sum_{i, j \neq i} \left(\frac{{p}^{2}_{i}}{2m} + \frac{4\pi a \hbar^{2}}{m}\delta(r_{ij})\right).
\end{equation}	
The delta interaction is the so called unitary interaction and is compensated by an overall wavefunction $\psi({r}_{1},...{r}_{N})$ satisfying the Bethe - Peierls boundary condition \cite{Blatt} 
\begin{equation}\label{e2}
	\lim_{r_{ij}\rightarrow 0}\frac{\partial \ln(r_{ij} \psi)}{\partial r_{ij}}=-\frac{1}{a}.
\end{equation}
In the physics of cold atoms \cite{Pethick} the trapping of particles into a potential translates the boundary condition of the unitary interactions to the form \cite{WC}
\begin{equation}\label{e21}
	\lim_{r_{ij}\rightarrow 0} \psi(R)= \frac{C}{r_{ij}}+ O(r_{ij}) - \frac{1}{a}.
\end{equation}
   Hyperspherical coordinates are used where $\psi(r_{1},\cdots, r_{N})=\psi(R)$ with $R^{2}=\sum_{i}r^{2}_{i}$. The unitary limit occurs in the case of infinite scattering length $a \rightarrow \infty$ and reflects a maximum value for the interaction strength between two particles in the s-wave channel. In particular, this means the s-wave scattering amplitude between two particles reaches the value $f_{k}=1/ik$ where $k$ is their relative momentum. The physics in this case is said to be universal in view of the absence of a characteristic length scale in the boundary condition. In nuclear physics, this limit resembles the nucleon-nucleon interaction with the short-range repulsion.

An interesting relation which reflects the unitary limit is the zero-energy solution of the Schrodinger equation \cite{Son}
\begin{equation}\label{e3}
	\sum_{i} \frac{\partial^{2}}{\partial r^{2}_{i}}\psi({r}_{1},...{r}_{N})=0,
\end{equation}
with the scaling behavior
\begin{equation}\label{e4}
	\psi({r}_{1},...{r}_{N})=R^{\nu} \psi(\Omega_{k}).
\end{equation} 
The scaling behavior (\ref{e4}) is of interest in relation to nuclear collective effects. Here, $\nu$ is a scaling exponent and $R$ has dimensions of length setting an overall scale for the distances between the particles within a trapping potential. The angles $\Omega_{k}$ reflect ratios between the $r_{i}$. For instance, in the case of $N=2$ particles the angle is $\Omega_{1} \equiv \alpha$ and defined by the ratio $r_{1}/r_{2}=\tan \alpha$. 
\section{An algebraic comparison}\label{2}
I would like now to discuss the similarity between the Schrodinger equation of the $U(6) \supset O(6)$ limit of the IBM \cite{Mexico}  and the non-relativistic Schrodinger equation for N particles in hyperspherical coordinates \cite{WC} confined in a harmonic oscillator trap. These two equations are
\begin{widetext}
\begin{equation}\label{Mex}
	- \frac{\hbar^{2}}{2M} \left(\frac{1}{\rho^{5}} \frac{\partial}{\partial \rho}\rho^{5} \frac{\partial}{\partial \rho} - \frac{ \sigma ( \sigma+ 4 ) }{\rho^{2}}\right)\Phi(\rho)+\frac{1}{2}M\omega^{2}\rho^{2} \Phi(\rho) = 
	\left(N_{b} + \frac{6}{2} \right)\hbar \omega \Phi(\rho),
\end{equation}
\begin{equation}\label{WC}
	-\frac{\hbar^{2}}{2M} \left(  \frac{1}{R^{3N-1}} \frac{\partial}{\partial R}R^{3N-1} \frac{\partial}{\partial R} -\frac{ \Lambda}{R^{2}}\right)\Psi(R) + 
	\frac{1}{2} M\omega^{2} R^{2}\Psi(R) =E \Psi(R),
\end{equation}
\end{widetext}
and mentioned below as the IBM equation (\ref{Mex}) and the hyperspherical equation (\ref{WC}) respectively. The solutions of these equations are radial wavefunctions and the similarity is discussed via the radial parts of the Schrodinger equations.

The unitary limit is introduced in the hyperspherical Eq (\ref{WC}) by setting $\omega=0$ and $E=0$ \cite{WC} with a specific state $\Psi(R)$ which satisfies the boundary condition (\ref{e21}) for $a \rightarrow \infty$. As is well known by the studies in cold atoms \cite{WC, Pitaevskii}, at any $N$ the trapping of the unitary scattered particles manifests the $SO(2,1)$ group. This group is isomorphic to the conformal group in one dimension \cite{Hagen} in which we deal with three generators that obey the commutation relations   
\begin{equation}\label{e6}
	[H,D]=-2i H, \quad [K,D]=2i K, \quad [K,H]=i \hbar^{2} \omega^{2} D.
\end{equation}
 $H$ is the free Hamiltonian of Eq (\ref{e1}) i.e the kinetic term without the unitary interaction which is absorbed into the boundary condition. In hyperspherical coordinates, $H$ is translated to the kinetic term of the hyperspherical equation (\ref{WC}). The harmonic trap of the same equation defines the special conformal operator $K=(1/2) M \omega^{2} R^{2}$ \cite{Mehen}. The dilatation operator is $D=-i R\partial_{R}+3N/2i$ with the eigenvalue of $R\partial_{R}$ to be denoted as $\nu$. The eigenvalue $\nu$ is the scaling exponent in Eq (\ref{e4}). The one dimensional conformal group consists of the aforementioned three generators and is larger than the group of simple scale transformations that are produced only by one generator, the dilatation operator $D$.
 
  The symmetry-based approach to the unitary limit \cite{WC} consists of the following two steps. First, a specific state of zero energy is defined in absence of the trap $(\omega=0)$ that obeys the scaling behavior of Eq (\ref{e4}) and respects the boundary condition (\ref{e21}) in the limit of infinite scattering length. This is the state at unitarity  denoted as $\psi^{0}_{\nu}$. Second, $\psi^{0}_{\nu}$ is mapped to states within the trap by the three generators of the $SO(2,1)$ group. This approach is applied in the IBM equation (\ref{Mex}) through the similarity of the latter with the hyperspherical equation (\ref{WC}) that is presented below.

The IBM equation (\ref{Mex}) is realized in the six dimensional space ($d=6$) of the $s$ and $\bf{d}$ bosons \cite{Mexico, Panos}. The boson number radius is $\rho=\sqrt{\beta^{2}+q_{0}^{2}}$. $\beta$ is the quadrupole deformation of the nuclear surface defined by $\beta^{2}=\sum_{i=1}^{5}q^{2}_{i}$. The five quadrupole coordinates $q_{i}$ define the five dimensional quadrupole plane where the $\bf{d}$ boson lives. The $s$ boson coordinate $q_{0}$ is a sixth transversal coordinate to this plane. 

On the other hand, the hyperspherical equation (\ref{WC}) is realized in standard hyperspherical coordinates. The number of dimensions  $d=3N$ is related with the number of particles in absence of the center of mass.

In the IBM equation (\ref{Mex}), the numerator of the centrifugal term is the eigenvalues $\sigma(\sigma+4)$ of the angular wavefunctions that span the irreducible representations of the $O(6)$ group. Now, the corresponding angular eigenvalues $\Lambda$ of the hyperspherical equation (\ref{WC}) obey the relation \cite{BEC}
\begin{equation}\label{e7}
	\Lambda^{2} Y_{\lambda \mu}(\Omega)=\lambda(\lambda+3N-2)Y_{\lambda \mu}(\Omega) \equiv \Lambda Y_{\lambda \mu}(\Omega).
\end{equation}
$\Lambda^{2}$ is the second order Casimir operator of the $O(d)$ orthogonal group in $d$ dimensions with $d=3N$. The zero energy solution (\ref{e3}) occurs for $\omega=0$ and is translated to the eigenvalues of the angular operator \cite{WC}
\begin{equation}\label{e9}
\Lambda=\nu (\nu+3N-2),
\end{equation}
which sets the scaling exponent $\nu$ in position of the $O(d)$ quantum number $\lambda$. Therefore in the zero energy solution $\lambda=\nu$ and the $O(d)$ group is preserved as a result of scale invariance.

In the hyperspherical equation the $O(6)$ group is revealed by setting $3N-2=4$ in the eigenvalues of Eq (\ref{e7}) that take the form $\lambda(\lambda+4)$. Namely, for $N=2$ the hyperspherical equation (\ref{WC}) lives in $d=6$ dimensions manifesting the $O(6)$ group. As a result, the radial part of the hyperspherical equation is formally similar with the radial part of the IBM equation. In both equations, the number of dimensions $d$ is present in the first term as the exponent of the radius $R^{d-1}$ and in the numerator of the centrifugal term via the eigenvalues of the generalized angular momentum for the angular wavefunctions. In $d$ dimensions, the algebraic form of the eigenvalues of the Casimir operator of the $O(d)$ group is $\lambda(\lambda+d-2)$ \cite{WC}. This algebraic form is invariant with respect to the decomposition of the $O(d)$ group in a specific chain of subgroups. The decomposition of the $O(6)$ group in the IBM is the non-canonical chain \cite{Iachello} $O(6) \supset O(5) \supset O(3) \supset O(2)$  while in the hyperspherical equation is the canonical chain \cite{Smirnov} $O(6) \supset O(5) \supset O(4) \supset O(3) \supset O(2)$. However, due to the invariance of the algebraic form of the eigenvalues of the $O(6)$ Casimir in both cases, a correspondence of the radial parts is algebraically valid from the correspondence of $\lambda$ with $\sigma$ i.e for $d=6$, $\lambda(\lambda+6-2) \rightarrow \sigma(\sigma+4)$. This correspondence permits to apply the symmetry-based approach to the unitary limit in the IBM equation and to define the dilatation operator, the special conformal operator and the free space Hamiltonian in the IBM via the same radial algebraic forms. The correspondence by no means implies a transfer of quantum numbers from the solutions of the IBM equation to those of the hyperspherical equation and vice-versa. The physical interpretation is related with the pair of wavefunctions that defines a reaction channel \cite{Blatt} as a projectile-target system and is presented below. The applicability of the IBM to reactions also has been discussed in previous works e.g in \cite{Alhassid}.

\subsection{Interpretation}\label{intr}
The algebraic comparison leads to the consideration of the generalised Hamiltonian
\begin{equation} \label{b1}
	H_{c}=H_{2N}+H_{IBM}+H_{2N/IBM},
\end{equation}
with $H_{2N}$ the Hamiltonian of even number of nucleons in hyperspherical coordinates, $H_{IBM}$ the IBM Hamiltonian in the $O(6)$ limit and a coupling term $H_{2N/IBM}$. The discussion from now on refers to a scattering process of $N=2$ incident nucleons onto the collective state of a target nucleus. In nuclear reactions, the potential of one incident nucleon on a target nucleus is usually taken to be the Coulomb plus a Woods-Saxon term. In order to study the unitary limit of $H_{c}$ like in the symmetry-based approach in cold atoms \cite{WC}, the general principle is to replace Woods-Saxon terms by appropriately defined unitary interactions. The $N=2$ incident nucleons are restricted to be neutrons and thus Coulomb interactions in $H_{2N}$ as well as in $H_{2N/IBM}$ are absent. 

\begin{figure}
	\includegraphics[scale=0.5]{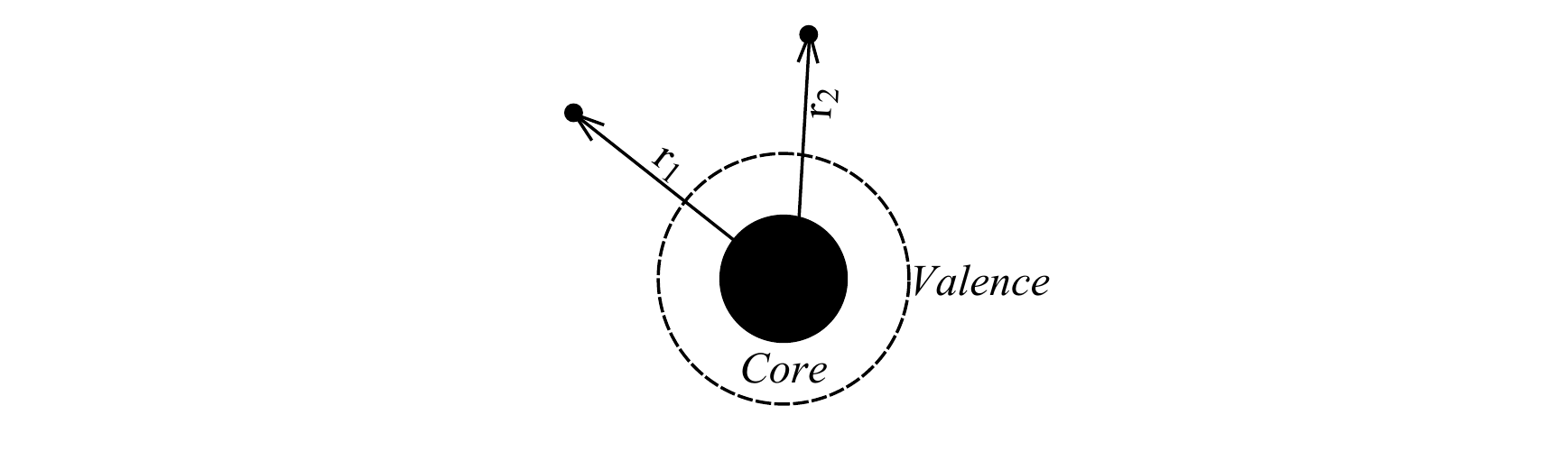}
	\caption{A snapshot that illustrates the relative positions $r_{1}$ and $r_{2}$ of the incident pair of neutrons 1 and 2 with respect to the core of the heavy and even-even target nucleus. The target nucleus is identified by the region of the core (circle filled with black color in the center) and the valence space (dashed circe homocentered with the core). The hyperspherical coordinate $R^{2}=r^{2}_{1}+r^{2}_{2}$ represents the relative coordinate between the incident pair and the core or the size of the compound system composed of the target nucleus plus two neutrons. The boson number radius $\rho$ represents the size of the valence space as shown by the dashed circle.}
	\label{f2}
\end{figure}

The positions $r_{1,2}$ of the incident neutrons are measured with respect to the core of the target nucleus with $R^{2}=r^{2}_{1}+r^{2}_{2}$  as shown in Figure \ref{f2}. The hyperspherical coordinate $R$ is expressed as $R^{2}=r^{2}_{12}/2+2R^{2}_{12}$, with $r_{12}=r_{1}-r_{2}$ and $R_{12}=(r_{1}+r_{2})/2$.

The choice of the core's position as the origin is an application of the more general case of the interaction of two similar particles with a fixed potential field represented by a {\it heavy} third body \cite{Morse}. Such an application has been realized for instance in excited electron pairs in atoms \cite{Fano}. In our case, that fixed potential field is represented by the collective nuclear state and each vector $r_{1,2}$ points directly to the mass of each similar particle here the neutron mass $M$. The choice of the core's position as the origin is indicated by the large mass of the {\it heavy} target nucleus with respect to the light mass of two incident neutrons. In nuclear structure, the binding energy of a heavy nucleus is large enough with respect to the binding energy of two nucleons - a difference of two orders of magnitude at least. In general, such a difference is evident in the binding energy of the deuteron. Particularly, in our case that difference is evident in the binding energy of two neutrons in a heavy nucleus i.e in the so called two-neutron separation energy ($S_{2n}$) with respect to the binding energy of the heavy nucleus.

 Each term of $H_{c}$ in (\ref{b1}) is in one to one correspondence with the form
\begin{equation}\label{b2}
	H_{c}=H(R)+ H (\rho)+H(\rho,R).
\end{equation}
 The kinetic term of $H(R)$ represents the relative kinetic energy between the incident neutrons and the target nucleus. For their scattering with respect to the collective state it is preferable to introduce the variable $r=R-\rho$. $R$ defines the position of the pair with respect to the core while $\rho$ indicates the position of the collective state. The kinetic energy is that of the hyperspherical equation (\ref{WC}) for $N=2$ with $r$ replacing $R$ denoted as $T_{r}$ with the mass $M$ to indicate the neutron mass.
 
 The potential that the two neutrons are subjected to in the vicinity of the collective state is absorbed into the coupling term $H(\rho,R)$. What remains to be determined in $H(R)$ is the interaction between the incident neutrons themselves. This is written in the form $g\delta(r_{1}-r_{2})$ with $g=4\pi \hbar^{2} a/M$ where $a$ is the scattering length of the neutron-neutron interaction. The zero-range approximation for the interaction betweeen the incident neutrons is due to the fact that internucleon separations in the valence space of a heavy nucleus are much larger than the range of the strong force. This situation is similar to that of dilute and cold atomic gases where interatomic separations are much larger than the range of  interatomic interactions \cite{Pethick}. This interaction is therefore compensated by the boundary condition of the type (\ref{e21}).

The states of $H_{c}$ are written as a sum over a pair of wavefunctions $\Psi(r,\rho)=\sum_{n}\Psi_{n}(r)\Phi_{n}(\rho)$ with $n$ to indicate a specific reaction channel. The states of the target (residual) nucleus $\Phi_{n}(\rho)$ are provided by the IBM equation while the states of the ingoing (outgoing) neutrons $\Psi_{n}(r)$ are provided by a hyperspherical equation of the type of Eq (\ref{WC}) for $N=2$ with $r$ replacing $R$. In what follows it is useful to apply the Feshbach formalism \cite{Feshbach}. Therefore consider the case of one open channel (they can be more but choose this for simplicity) with $P=|\Phi_{0}(\rho)\rangle \langle \Phi_{0}(\rho)|$ and a set of closed channels with $Q=\sum_{n>0}|\Phi_{n}(\rho)\rangle \langle \Phi_{n}(\rho)|$. For instance the state of the open channel is $P\Psi(r,\rho)=\Psi_{0}(r)|\Phi_{0}(\rho)\rangle$ and consists of the amplitude $\Psi_{0}(r)$ in $R$ space in the "direction" of the vector $|\Phi_{0}(\rho)\rangle$ in $\rho$ space. Elastic processes are those in which the scattering state involves only the open channel. Inelastic processes are those where the channel changes during the reaction and the scattering state involves components both from $P$ and $Q$ sets.

The term $H(\rho)$ is the Hamiltonian of the IBM equation. $H(\rho)$ applies to the wavefunction $\Phi_{n}(\rho)$ that represents the amplitude of the IBM state $|N_{b},\sigma,\tau,L \rangle$. Apart from the boson number $N_{b}$, the collective state is characterized by the $O(6)$ quantum number $\sigma$, the $O(5)$ quantum number $\tau$ that labels target states with respect to $\gamma$ unstable rotations and the angular momentum $L$. In the channel formalism of nuclear reactions, the Hamiltonian of the reaction is integrated over the target states. In our case this Hamiltonian is $H_{c}$. The "internal" collective coordinate of the target nucleus is $\rho$ and serves as the radial integration variable that feeds the valence space. The former term $H(R)$ is insensitive to these integrations as they are realized over the wavefunctions of the IBM equation. However, $H(\rho)$ contributes to the energy by the term
\begin{equation}\label{b21}
\langle L,\tau,\sigma,N_{b}| H(\rho)|N_{b},\sigma,\tau,L \rangle.
\end{equation}
For instance, the entrance channel ($n=0$) is chosen to be that one where the target is on its ground state with $L=0$, $\tau=0$ and $\sigma=N_{b}$. The contribution of $H(\rho)$ is then $E_{0}$ which is the energy of the ground state of the target nucleus. The excited channels refer to target states with $N_{b}+1$, $N_{b}+2$ bosons and so on. The same type of calculations is then performed for $n=1,2 \cdots$ and the contribution of $H(\rho)$  is written as $E_{1}, E_{2}$ and so on. From all the substates of each excited channel, we examine target states with $L=0$, $\tau=0$ and $\sigma=N_{b}+1,2, \cdots$ for $n=1,2, \cdots$ respectively.

 The coupling term $H(\rho,R)$ reflects the potential the two neutrons see by entering the vicinity of the collective state of the target nucleus. For low energies of the incident neutrons, lower than single particle excitations of the target nucleus, the examination of a pair-collective state resonance leads to the analysis of the scattering in terms of a pair-collective state coupling. The pair is explicitly defined by its relative position $r$, conjugate momentum $k_{r}$ with respect to the collective state and the angular momentum of an incident $s$-wave. Such an analysis occurs also as the outcome of the $O(6)$ symmetry which indicates a coupling of the incident pair as a whole with the collective state. That coupling is maximum when the size of the coupled system as indicated by $R$ coincides with the boson radius $\rho$ of the collective state i.e when $r=0$. Therefore, in the approximation of interactions of zero-range in the valence space, the pair-collective state coupling takes the form  $H(\rho,R)=g_{r}\delta(R-\rho)$. The coupling constant $g_{r}$ is discussed in subsection \ref{uni} and remarks in relation with molecular states of cold atoms are recorded in subsection \ref{5}  .
 
 For the application of $H_{c}$ to a particular type of reaction one needs to add a second extra term in the pair-collective state coupling $H(\rho,R)$ that accounts for changes in the boson number of target states. The motivation of such a term is found in the Hamiltonian used in trapped cold atoms \cite{Timmermans} where an external magnetic field creates the open - closed channel coupling and not the unitary interaction which is absorbed in the scattering wavefunction. In the case of two ingoing (outgoing) neutrons, the role of the external field is reflected on the impact of the incident (outgoing) pair on the IBM target states as the addition (removal) of one boson. Therefore, the coupling is separated into two terms as $H(\rho,R)=g_{r}\delta(r)+F$, with $\delta(r)$ to affect the pair state in the relative coordinate $r$ and $F$ the extra term that changes the target (IBM) states. For the particular case of a two neutron transfer the extra term is written as $F=s^{\dagger}+s$. In such a case $F$ represents the arrival (departure) of the pair as the addition (subtraction) of one boson in the IBM basis states which constitute the target states of $H_{c}$.

  In the rest of this subsection attention is paid on the scattering of the pair with the collective state. Detailed calculations of all the results quoted below are presented in Appendix \ref{app}.

  The pair-collective state coupling $H(\rho,R)$ gives rise to a pair-collective state scattering length $a_{r}$ that is defined by the equation
\begin{equation}\label{b3}
	k_{r}\cot\delta_{0}=-\frac{1}{a_{r}(k_{r})},
\end{equation}
with $k_{r}$ the conjugate wavenumber with respect to $R-\rho$. $a_{r}(k_{r})$ is a generalized scattering length in the sense of \cite{Bethe}. $\delta_{0}$ represents the $s$-wave phase shift of the incident pair between the ingoing ($k^{-}_{r}$) and the outgoing wave ($k^{+}_{r}$). This $s$-wave corresponds to a total angular momentum zero of the incident neutrons with respect to the core i.e $l_{1}+l_{2}=0$. Namely, the examination of the pair-collective state scattering is restricted to the $l=0$ component of the scattering cross section. To this end what follows refers to the coupling of the incident $s$-wave of the pair of neutrons with the collective $0^{+}$ ground state of the target nucleus. 

The scattering state may be written in the form of an incident plane wave plus a spherical wave at large distances from the target as
\begin{equation}\label{b4}
	\lim_{r\rightarrow \infty}\Psi_{0}(r)=e^{ik^{-}_{r}r}+\frac{f_{k_{r}}}{r^{5/2}}e^{ik^{+}_{r}r}.
\end{equation}
This form is indicated by the $r^{-(d-1)/2}$ factors of spherical waves in hyperspherical coordinates in $d=6$ dimensions where the solid angle is $\pi^{3}$ \cite{Morse}. In this geometry one derives scattering amplitudes and reaction cross sections by following the same steps as in ordinary scattering but with the partial wave expansion to be written in terms of the $\lambda$ quantum number of the $O(6)$ symmetry. The result for the cross section $\sigma$ and the scattering amplitude $f_{k_{r}}$ of the $s$-wave ($\lambda=0$) is
\begin{equation}\label{b6}
\begin{split}
	& \sigma=(4\pi)^{3}\frac{\sin^{2}\delta_{0}}{k^{2}_{r}}= \frac{(4\pi)^{3}}{k^{2}_{r}} \left(\frac{1}{1+\cot^{2}\delta_{0}} \right)= \\ & \frac{(4\pi)^{3}}{1/a_{r}^{2}(k_{r})+k^{2}_{r}}, \quad f_{k_{r}}=\frac{8}{1/a_{r}(k_{r})-ik_{r}}.
		\end{split}
\end{equation}
The maximum value of the scattering amplitude $|f_{k_{r}}|\leq 8/k_{r}$ occurs for  $1/a_{r}(k_{r})=0$. This condition, the zero value of $1/a_{r}(k_{r})$, defines the resonance of the incident pair of neutrons with the target nucleus.
\subsubsection{Effective range}\label{eff}
 For low $k_{r}$,  $1/a_{r}(k_{r})$ is expanded in the form
\begin{equation} \label{b600}
\frac{1}{a_{r}(k_{r})}=\frac{1}{a_{r}}-\frac{1}{2}k_{r}^{2}r^{*} + \cdots.
\end{equation}
Such expansions are used in the effective range approximation for proton-neutron scattering \cite{Blatt}, in resonances of a single neutron on a target nucleus \cite{Bethe} but also in resonances between cold atoms \cite{Pricoupenko}. In our case, the effective range
$r^{*}$ represents a characteristic length of the interaction between the incident pair and the collective state. The magnitude of $r^{*}$ is related with the width of the resonance formed by the incident pair with the collective state as $\Gamma= \hbar^{2} k_{r}/(M r^{*})$. The condition for the realization of such a resonance in the heavy compound nucleus is $1/a_{r}(k_{r})=0$ \cite{Bethe}. 
\subsection{Intermediate states}\label{solutions}
Feshbach resonances in systems of cold atoms exemplify the occurence of intermediate states of the Feshbach formalism \cite{Timmermans}. Intermediate states were originally proposed to occur in compound nuclei \cite{Feshbach} and may constitute a candidate for the unitary limit in heavy nuclei in  analogy with the same limit of cold atoms. To this end, one considers now coupled channels equations in which energies, of the open as well as of the closed channels, are measured with respect to the ground state energy $E_{0}$ of the target nucleus. 
In general channel-channel couplings like $H_{PQ}$ or $H_{QQ}$ consist of matrix elements $\langle \Phi_{n'}(\rho)|H(\rho,R)|\Phi_{n}(\rho) \rangle$ as integrals of the form 
\begin{equation}\label{b71}
\int d\rho \rho^{5}\Phi_{n'}(\rho)H(\rho,R) \Phi_{n}(\rho) \equiv H_{nn'}.
\end{equation}
For instance this is $H_{00}$ in the open channel, $H_{11}$ in the first closed channel  and $H_{01}$ is their coupling.
The Hamiltonian $H_{QQ}=QHQ$ is a matrix of the whole vector space composed of the target states with $N_{b}+1$, $N_{b}+2$ bosons and so on. By writing the total energy of the target plus two neutrons as $E$, the wavefunction of the pair is provided by the solutions of the coupled channels equations
\begin{equation}
\begin{split}\label{e26}
(E-H_{PP})P|\Psi \rangle=H_{PQ} Q |\Psi \rangle,\\
(E-H_{QQ})Q|\Psi \rangle=H_{QP} P|\Psi \rangle.
\end{split}
\end{equation}
In general $|\Psi \rangle= \sum_{n} |\Psi_{n}\rangle \otimes |\Phi_{n}(\rho)\rangle$ with the tensor product to indicate the coupling of the total angular momentum of the incident pair with the total angular momentum of the collective state. The angular momentum coupling of the incident $s$ wave with the $0^{+}$ ground state of the target nucleus is scalar and the sum is reduced to the form $|\Psi \rangle= \sum_{n} |\Psi_{n}\rangle |\Phi_{n}(\rho)\rangle$. The amplitude of the pair is $ \langle r|\Psi_{0} \rangle= \Psi_{0}(r)$ in the open channel. This amplitude respects the boundary condition $\lim_{r_{1}\rightarrow r_{2}}\Psi_{0}(r)$ of the type (\ref{e21}) which reflects the unitary interaction between the incident neutrons themselves of coupling $g$. 

 A closed channel state is defined as $|c_{n} \rangle= |\Psi_{n} \rangle |\Phi_{n}(\rho) \rangle $ with $n > 0$ while $|c_{0} \rangle = |\Psi_{0} \rangle |\Phi_{0}(\rho) \rangle$ defines the open channel. A second boundary condition applies to the amplitude $\langle r|c_{0} \rangle=\Psi_{0}(r,\rho)$ as $\lim_{r \rightarrow 0 }\Psi_{0}(r,\rho)$ reflecting the pair-collective state unitary interaction of coupling $g_{r}$. 
 
 In Eq (\ref{e26}), the open-closed channel couplings $H_{QP}$ and $H_{PQ}$ are calculated from the coupling term $H(\rho,R)=g_{r}\delta(r)+F$. The unitary part of the coupling is absorbed in the wavefunction $\Psi_{0}(r,\rho)$ by the second boundary condition. The remaining part $F=s^{\dagger}+s$ changes the boson number of the target states. Therefore, the couplings between the open channel of $N_{b}$ bosons with the first closed channel of $N_{b}+1$ bosons are $H_{10}=\langle N_{b}+1|s^{\dagger}|N_{b}\rangle=\sqrt{N_{b}+1}$ and $H_{01}=\langle N_{b}|s|N_{b}+1\rangle=\sqrt{N_{b}+1}$. The $F$ term is a two-neutron transfer operator for the compound nucleus and does not contribute to the open-open $H_{PP}$ and closed-closed $H_{QQ}$ channel couplings.
\subsubsection{Open channel state}
 From the first equation of (\ref{e26}) and for zero couplings between different channels, one starts with the homogeneous equation of the open channel
\begin{equation}\label{e29}
(E-H_{PP})\Psi^{+}_{0}(r)|\Phi_{0}(\rho) \rangle=0.
\end{equation}
The ket of the IBM state $|\Phi_{0}(\rho) \rangle$ is present indicating the relevant entry in the matrix for the coupled equations. The solutions of this equation determine the asymptotic behavior of the pair wavefunction $\Psi_{0}(r)$ denoted as $\Psi^{+}_{0}(r)$ which obeys the scattering condition (\ref{b4}). $\Psi^{+}_{0}(r)$ is analyzed in partial waves of $O(6)$ symmetry as presented in detail in Appendix \ref{app}. The Hamiltonian of the open channel reads 
\begin{equation}\label{e28}
H_{PP}=T_{r}+\langle \Phi_{0}(\rho)|H(\rho)| \Phi_{0}(\rho)\rangle+\langle  \Phi_{0}(\rho)|H(\rho,R)| \Phi_{0}(\rho)\rangle.
\end{equation}
The coupling $\langle  \Phi_{0}(\rho)|H(\rho,R)| \Phi_{0}(\rho)\rangle \equiv H_{00}$ is absorbed into the wavefunction $\Psi^{+}_{0}(r)$. This absorption results from the second boundary condition $\lim_{r \rightarrow 0}\Psi^{+}_{0}(r)$. The mean value of the IBM Hamiltonian $\langle \Phi_{0}(\rho)|H(\rho)| \Phi_{0}(\rho)\rangle=E_{0}$ is zero in the chosen scale. Therefore, from the Hamiltonian $H_{PP}$ of the open channel, only the kinetic term $T_{r}$ survives explicitly in Eq (\ref{e29}). This equation is of the same form with the hyperspherical equation (\ref{WC}) for $N=2$ particles without the trap and $r$ replacing $R$. The substitution of $\Lambda=\lambda(\lambda+4)$ in the numerator of the centrifugal term gives
\begin{equation}\label{e280}
	-\left(  \frac{1}{r^{5}} \frac{\partial}{\partial r}r^{5} \frac{\partial}{\partial r} - \frac{\lambda(\lambda+4)}{r^{2}} \right)\Psi^{+}_{0}(r)=k^{2}_{r} \Psi^{+}_{0}(r),
\end{equation}
with $k^{2}_{r}=2 M E/\hbar^{2}$. As analyzed in Appendix \ref{app0}, this equation gives a spherical wave of the form $\Psi^{+}_{0}(r)=r^{-5/2}\chi(r)$ or a Bessel function of the form $r^{-2}\chi_{\lambda+2}(k_{r}r)$ that converges to the spherical wave far away from the reaction center. The general form of the asymptotic solution for the $s$-wave ($\lambda=0$)  reads
\begin{equation}\label{e281}
\Psi^{+}_{0}(r)=8 e^{i \delta_{0}} \frac{\chi_{0+2}(r)}{k_{r}r^{5/2}}P_{0}(\cos \alpha |4).
\end{equation}
The angular part is determined by the Gegenbauer polynomial $P_{0}(\cos \alpha |4)$ over the angles $\alpha$ and $4$ which indicates the azimuthal and polar angles $(\theta_{1},\phi_{1})$ and $(\theta_{2},\phi_{2})$ of each neutron. For the $s$-wave, the zero order makes the Gegenbauer polynomial unity. The numerical factor of $8$ results from the $\lambda=0$ component of the partial wave expansion.
 
 The boundary condition for $r \rightarrow 0$ determines the radial part of the solution i.e the kind of the Bessel function as presented in detail in Appendix \ref{appBC}. Two cases of solutions are obtained. In the first one, the incident pair interacts with the collective state as a hard sphere of radius $a_{r}$ and the solution near the reaction center is
\begin{equation}\label{e2821}
u_{0}(r)=C\left(\frac{N_{0+2}(k_{r}r)}{r^{2}}+\frac{4}{\pi k_{r}^{2}a^{4}_{r}}\right),
\end{equation}
with $C=\sqrt{\pi/(2k_{r})}e^{i5\pi/4}$ and $N_{0+2}(k_{r}r)$ the Neummann function of order $0+2$ for the $s$-wave.  In this case, the channel wavefunction $\Psi_{0}(\rho,r)=\Phi_{0}(\rho)u_{0}(r)$ satisfies the boundary condition
\begin{equation}\label{e282}
\lim_{r\rightarrow 0}\Psi_{0}(\rho,r)=\Phi_{0}(\rho)\left(\frac{1}{r^{4}}-\frac{1}{a^{4}_{r}} \right).
\end{equation} 
The boundary condition (\ref{e282}) compensates the unitary pair-collective state interaction in the so called shape-elastic scattering i.e when the entrance channel does not change at the eve of the scattering.

In the second case, the hard sphere approximation is removed. This occurs when the range of the interaction $r^{*}$ is smaller than $a_{r}$. The de Broglie wavelength of the relative motion is $1/k_{r}$ and the region of interest is $r^{*}<r<k^{-1}_{r}$. This region is beyond the range of the pair-collective state interaction and defines the vicinity of the collective state relative to the scattering of the incident pair. For that region the solutions are normalized to the wavefunction
\begin{equation}\label{e283}
u_{0}^{+}(r)=\frac{8i^{5/2}}{k_{r}r^{5/2}}\sin\left(k_{r}r-\frac{5\pi}{4}+\delta_{0}\right).
\end{equation}
 The pair wavefunction of the open channel far away from the reaction center is $\Psi^{+}_{0}(r)=e^{i\delta_{0}}u_{0}^{+}(r)$, where $\Psi^{+}_{0}(r)=\lim_{r \rightarrow \infty} \Psi_{0}(r)$. The wavefunction $\Psi_{0}(r)$ is written in terms of an incident Neummann plus an outgoing Hankel function
 \begin{equation}\label{B812}
 \Psi_{0}(r)=8C \left( \frac{ N_{0+2}(k_{r}r)}{r^{2}}+\frac{(e^{2i\delta_{0}}-1)}{2ik_{r}}\frac{H^{(1)}_{0+2}(k_{r}r)}{r^{2}} \right).
 \end{equation} 
 In this case the boundary condition is
  \begin{equation}\label{B81d}
 \lim_{r \rightarrow 0}\Psi_{0}(\rho,r)=\Phi_{0}(\rho)\frac{(-1+ia_{r})}{r^{4}}.
 \end{equation}
which resembles an optical potential for the unitary pair-collective state coupling. In what follows, the scattering wavefunction is normalised in two isotropic waves in six dimensions 
\begin{equation}\label{e2914a}
\Psi^{+}_{0}(r)=8\frac{e^{-i(k_{r}r-5\pi/4)}}{r^{5/2}}-S_{0}8\frac{e^{ik_{r}r}}{r^{5/2}},
\end{equation}
with $\Psi^{+}_{0}(r)=-2i k_{r}e^{i\delta_{0}}u^{+}_{0}(r)$ and $S_{0}=e^{2i\delta_{0}}$.

\subsubsection{Coupled channels solutions}
 Having defined the asymptotic form of $\Psi^{+}_{0}(r)$ for zero couplings with the closed channels, there is a standard method \cite{Feshbach, Timmermans} to obtain the open channel solution of the coupled equations (\ref{e26}) through the intermediate states. These states are solutions of the homogeneous equation of the closed channel space,
\begin{equation}\label{b10}
(\epsilon_{n}-H_{QQ})\Psi_{n}(r) |\Phi_{n}(\rho) \rangle=0, \quad n>0,
\end{equation}
that results from the second equation of (\ref{e26}) for zero couplings with the open channel.

 Intermediate states represent {\it stationary} states of the $A+2n$ compound nucleus formed by the target plus two neutrons. They serve as resonance states of energy $\epsilon_{n}$ whith respect to the total energy $E$ of the open channel. From all these states, the amplitude of a specific intermediate state $\langle r|c_{m} \rangle = \Psi_{m}(r)| \Phi_{m}(\rho) \rangle$, $m>0$ is subjected to the $m$-th entry of the Hamiltonian matrix $H_{QQ}$ 
\begin{equation}\label{b10a}
	T_{r}+\langle \Phi_{m}(\rho)|H(\rho)| \Phi_{m}(\rho)\rangle+H_{mm}.
\end{equation} 
$H_{mm}$ stands for the pair-collective state unitary interaction in the closed channel and is absorbed in the channel wavefunction $\Psi_{m}(\rho,r)$ by the second boundary condition for $r\rightarrow 0$. That boundary condition is further analyzed in subsection \ref{symm} in Eq (\ref{c10b}). 

A specific intermediate state $m$ with energy very close to the energy of the open channel, satisfies the homogeneous equation
 \begin{equation}\label{e331a}
 (T_{r}+E_{m})\Psi_{m}(r)|\Phi_{m}(\rho)\rangle=\epsilon_{m}\Psi_{m}(r)|\Phi_{m}(\rho)\rangle.
 \end{equation}
Attention must be paid on the notation. The energy of the intermediate state is $\epsilon_{m}$ while the energy of the IBM state is $\langle \Phi_{m}(\rho)|H(\rho)| \Phi_{m}(\rho)\rangle = E_{m}$.
  The derivation of the width of the intermediate state is presented in detail in Appendix \ref{app11} along with all the results quoted in the rest of this subsection. 

 When the total energy $E$ of the open channel is close to the energy of the intermediate state $\epsilon_{m}$, the wavefunction of the open channel of Eq (\ref{e2914a}) takes the form
 \begin{equation}\label{e322a}
\begin{split}
& \Psi_{0}(r)=8i^{5/2}\frac{e^{-ik_{r}r}}{r^{5/2}}-  \left(1-\frac{i \Gamma_{m}}{E-E_{m'}+i\frac{\Gamma_{m}}{2}}\right)e^{2i\delta_{0}}8\frac{e^{ik_{r}r}}{r^{5/2}}.
\end{split}
\end{equation}
This is a wavefunction with a resonance for the scattering part of width
\begin{equation}\label{e3400a}
\Gamma_{m}=b^{2} \left(\frac{4 M} {\hbar^{2}}\right) k_{r} ,
\end{equation}
with the coupling $b \equiv \int dr \Psi_{m}(r)H_{m0}u^{+}_{0}(r)$. Since the target states of the reaction channels are defined by the boson number, the coupling to the first closed channel is $H_{m0}=\sqrt{N_{b}+1}$ and the quantity $b^{2}$ takes the simple form
\begin{equation}\label{e3400b}
b^{2}=(N_{b}+1){\bigg\rvert}\int dr \Psi_{m}(r)u^{+}_{0}(r){\bigg\rvert}^{2}.	
\end{equation}
The width of the resonance depends on the boson number times the probability amplitude of the overlap between the pair wavefunction of the open channel with the intermediate state. This overlap is a spectroscopic factor for the intermediate state.

 The energy of the resonance $E_{m'}=\epsilon_{m}+\Delta_{m}$ is the energy of the intermediate state $\epsilon_{m}$ shifted by the amount of $\Delta_{m}$ which also depends on the open-closed channel coupling.

The width of the resonance of the intermediate state gives rise to the scattering matrix  
 \begin{equation}\label{20}
	S_{0}=e^{2ik_{r}a_{r}}\left(1-i\frac{\Gamma_{m}}{E-E_{m'}+i\Gamma_{m}/2} \right).
\end{equation}
In the physics of compound nuclei the second term is named as a fluctuation to the cross section caused by the intermediate state \cite{Feshbach,Weidenmuller}. Such a fluctuation is the experimental observable of the intermediate state and signifies the compound-elastic reaction. In our case, that fluctuation tunes the pair-collective state scattering length. Namely, the width of the intermediate state defines the effective scattering length
\begin{equation}\label{21}
a_{reff}=a_{r}+\frac{1}{2k_{r}}\tan^{-1}\left(\frac{\Gamma_{m}(E-E_{m'})}{(E-E_{m'})^{2}+\Gamma_{m}^{2}/4} \right).
\end{equation}
\subsection{Unitary limit in heavy nuclei} \label{uni}
In the effective scattering length (\ref{21}) one studies the low energy limit when the total energy $E$ is very close to the energy of the resonance $E_{m'}$. In molecules that difference $E-E_{m'}$ refers to the detuning of the Feshbach resonance \cite{Timmermans}. One therefore has to determine the energy of the resonance. This is equivalent with the determination of the energy $E_{m'}$ of the intermediate state.

  The relation of the energy of the intermediate state with the energy of the IBM (target) state is clarified by the investigation of equation (\ref{e331a}). When the energy of the intermediate state is the IBM energy i.e when $\epsilon_{m}=E_{m}$, equation (\ref{e331a}) takes the form
\begin{equation}\label{22}
T_{r}\Psi_{m}(r)|\Phi_{m}(\rho)\rangle=0 \Psi_{m}(r)|\Phi_{m}(\rho)\rangle.
\end{equation} 
This is a hyperspherical equation with a zero energy solution that satisfies the condition of zero frequency of the trap. The channel wavefunction $\Psi_{m}(\rho,r)$ absorbs the unitary coupling $H_{mm}$ by the boundary condition. These are the conditions for the unitary limit in the sense of the hyperspherical equation (\ref{WC}). They are revealed by the condition $\epsilon_{m}=E_{m}$ which is called as the unitary condition below and in subsection \ref{5}.

Equation (\ref{22}) means that the energy of the pair in the $m$-th intermediate state of the compound nucleus is merely taken as the energy of the IBM state with $N_{b}+m$ bosons. Namely, pair states in closed channels which have the energy of the correspondent IBM state of specific rotational $L=0$, $\tau=0$ and vibrational $\sigma=N_{b}+m$ quantum numbers, play the same role with trapped-molecular states in systems of cold atoms. 

One examines now the behavior of the effective scattering length at the resonance. The effective scattering length goes to infinity when the quantity $E-(E_{m}+\Delta_{m})$ is close to zero. This is evident by Eq (\ref{21}) which for a low detuning $E-E_{m'}$ and in the limit of a very low energy of the incident beam $k_{r}\rightarrow 0$,  gives
\begin{equation}\label{23}
\lim_{E \rightarrow E_{m'}}a_{reff}=a_{r}-\frac{\Gamma_{m}}{2k_{r}(E-E_{m'})}.
\end{equation}
 By using the relation for the width of the intermediate state (\ref{e3400a}) one obtains the effective scattering length in terms of the neutron mass $M$ and the open-closed channel coupling $b$. The result is
\begin{equation}\label{24}
\lim_{E \rightarrow E_{m'}}a_{reff}=a_{r}-\frac{2M b^{2}}{\hbar^{2}(E-E_{m'})}.
\end{equation}
The last two relations indicate the tuning of the pair-collective state scattering length reaching its maximum value at the eve of the resonance with the intermediate state.

It is of interest now to explore the maximization of the effective scattering length in relevance to the energy independent scattering length $a_{r}$. At resonance, the inverse generalized scattering length is $1/a_{r}(k_{r})=0$ \cite{Bethe}. Close to that value, the generalized scattering length subjects to the effective range expansion 
\begin{equation} \label{B131}
\frac{1}{a_{r}(k_{r})}=\frac{1}{a_{r}}-\frac{1}{2}k_{r}^{2}r^{*} + \cdots.
\end{equation}
For $k_{r}\rightarrow 0$, the term $-(1/2)k^{2}_{r}r^{*}$ and higher order terms go to zero that is $k_{r} r^{*}<<1$. This is the case where $1/a_{r}(k_{r}) \rightarrow 1/a_{r}$ and therefore the effective scattering length $a_{reff}$ refers to the energy independent part $1/a_{r}$ of the expansion. 

The simple case of two coupled channels exemplifies this situation. With the unitary parts of the couplings $H_{00}$ and $H_{11}$ to be absorbed into the channel wavefunctions $\Psi_{0}(\rho,r)$ and $\Psi_{1}(\rho,r)$ respectively, the coupled equations of the open with the first closed channel read
\begin{equation}\label{b8}
	\begin{split}
		& (E-T_{r})\Psi_{0}(r)=H_{01}\Psi_{1}(r),\\
		& (E-T_{r}-E_{1})\Psi_{1}(r)=H_{10}\Psi_{0}(r).
	\end{split}
\end{equation}
 $E$ is the total energy of the open channel. The eigenvalue of $T_{r}$ is $\hbar^{2}k^{2}_{r}/2M$ and in the zero energy limit of the incident beam goes to zero as $k_{r} \rightarrow 0$. 

\begin{figure}
	\includegraphics[scale=0.5]{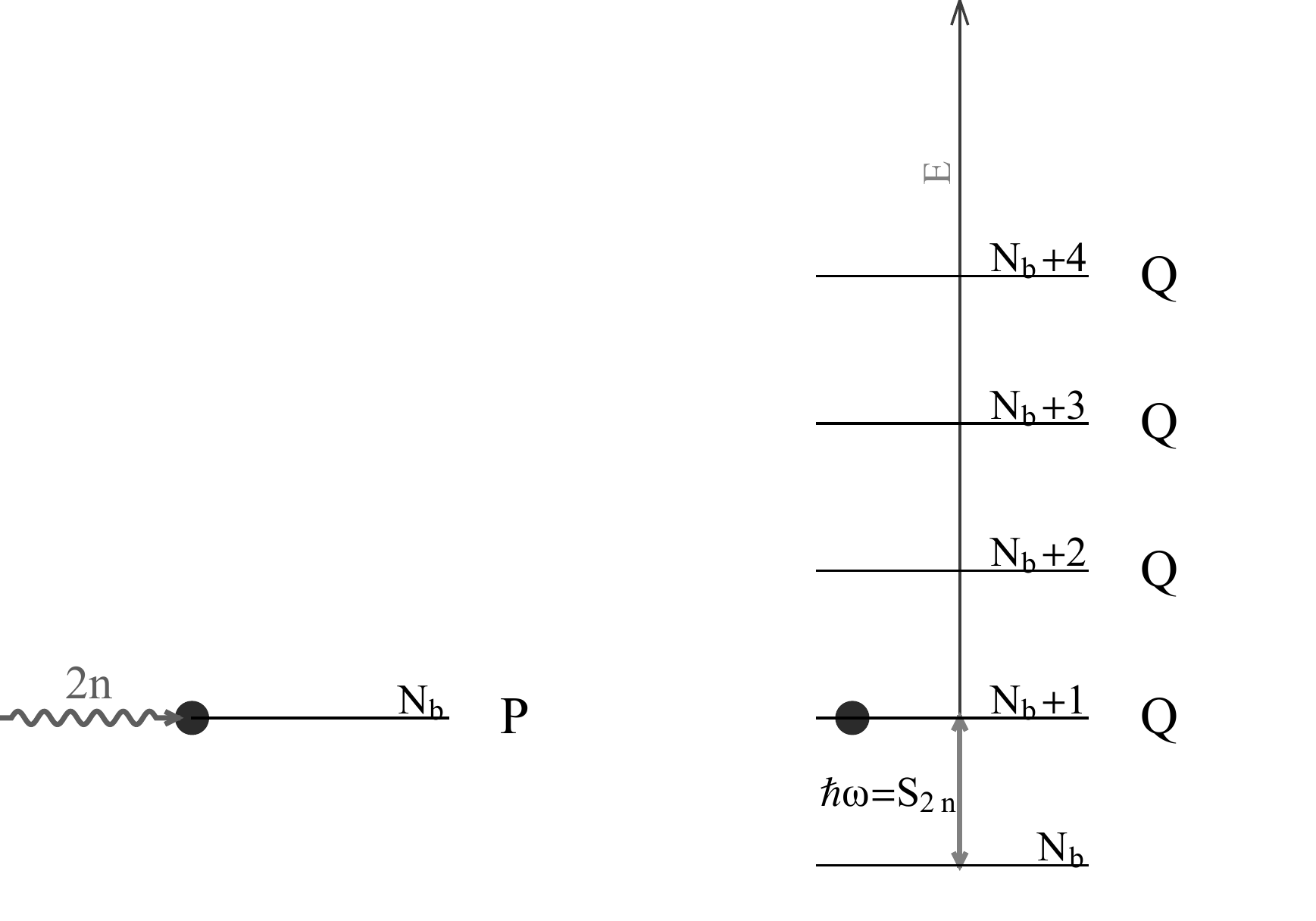}
	\caption{ Left: The open channel (P) consists of the incident wave of two neutrons  as they are represented by the squiggle arrow $2n$ and the ground state of the target nucleus as represented by the $N_{b}$ level of the IBM. In the IBM, the coupling of a pair of neutrons with the $N_{b}$ level raises the energy of the open channel to the $N_{b}+1$ level. Right: The set of resonances with the closed channels (Q) are provided by identifying the stationary (intermediate) states of the 
		$A+2n$ compound nucleus as IBM eigenstates. In analogy with cold atoms, the intermediate states are trapped states with the role of the trap to be provided by the IBM Hamiltonian. The energy difference between the trapped states is the two neutron separation energy $\hbar \omega=S_{2n}$.
		Remarks: In the Hamiltonian $H_{c}$, the open-closed and the closed-open coupling is provided by the term $s^{\dagger}+s$ which applies this scheme to a two-neutron transfer reaction. Such a process may be tested in an exotic $A+2n$ isotope which is short lived and therefore expected to decay back the $2n$ in a relatively short time. The process is experimentally testable as a fluctuation of specific energy and width to the cross section. This fluctuation tunes the pair-collective state scattering length.}	\label{f12}
\end{figure}

The equation of the intermediate state of the 1st closed channel is obtained by the second equation of (\ref{b8}) by setting  $H_{10}=0$. This choice restricts the total energy $E$ to the energy $\epsilon_{1}$ of the intermediate state i.e
\begin{equation}\label{b9}
	(\epsilon_{1}-T_{r}-E_{1})\Psi_{1}(r)=0.
\end{equation}
 The condition $\epsilon_{1}=E_{1}$ gives a zero energy state. By the onset of the coupling $H_{01}$ that zero energy state appears as a resonance in the open channel provided that the total energy $E$ of the open channel crosses with $E_{1}$. The crossing of the energies of the open and the first closed channel is illustrated in Figure \ref{f12} and explained in the subsection \ref{cs}.

That resonance tunes the scattering length by the amount of the open-closed channel coupling $b$ in the expression
\begin{equation}\label{B23a}
	\lim_{E \rightarrow E_{1'}}a_{reff}=a_{r}-\frac{2M b^{2}}{\hbar^{2}(E-E_{1'})}.
\end{equation}
The maximization of the effective scattering length occurs for $E=E_{1'}$ and is testable via a fluctuation of the cross section of determined energy and width as explained in detail in subsection \ref{mts}.
\subsubsection{Effective coupling constant}
In general, the unitary limit represents the case of maximum interaction strength of all the couplings involved in the Hamiltonian of the problem. The interaction strengths of $H_{c}$ are the following. The one between the incident neutrons themselves denoted as $g$ and the one between the incident pair with the collective state denoted as $g_{r}$. 
With respect to $g_{r}$, the coupling $H(\rho,R)=g_{r}\delta(r)+F$ determines the pair-collective state resonance by the open-closed channel couplings $H_{PQ}$ which are determined by the $F$ term. On their turn these couplings manifest the intermediate state which tunes the scattering length. It is shown immediately below how the terms that tune the scattering length are absorbed into the unitary coupling $g_{r}$ and define an effective coupling constant $g_{reff}$ in a similar sense with the definition of the effective coupling constant in the case of cold atoms.

The Born approximation gives also a type of relation for the effective scattering length \cite{Pethick}. For the coupling of the open channel with the $m=1$ closed channel this relation reads
\begin{equation}\label{b11}
\frac{4\pi^{3}\hbar^{2}}{M}a_{reff}=\frac{4\pi^{3}\hbar^{2}}{M}a_{r}+\frac{|\langle c_{1}|H_{01}|c_{0}\rangle|^{2}}{E-E_{1'}},
\end{equation}
with $a_{r}$ to be the scattering length of the open channel in absence of couplings. However, the Born approximation is not strictly valid in compound nuclei \cite{Bethe}. The general relation which survives beyond the Born approximation is that of the effective scattering length (\ref{21}). Nevertheless, in both cases the maximization of the scatttering length depends on the open-closed channel coupling and on the denominator $E-E_{1'}$. The effective scattering length $a_{reff}$ becomes maximum when the total energy of the open channel $E$ crosses with the energy of $ E_{1}=N_{b}+1$ bosons up to the shift $\Delta_{1}$.

A similar situation arises in systems of cold atoms where the Born approximation cannot be used for low-energy binary atom collisions  \cite{Timmermans}. In such a case, the interaction strength is replaced by the effective interaction strength i.e $g_{r}$ goes to $g_{reff}$. In absence of resonance, the coupling $g_{r}$ is associated with the energy independent $a_{r}$ as $g_{r}=4\pi^{3}\hbar^{2}a_{r}/M$. The effective coupling constant in the presence of the pair-collective state resonance is
\begin{equation}\label{B24}
g_{reff}=\frac{4\pi^{3}\hbar^{2}}{M}a_{reff}=g_{r}-\frac{8 \pi^{3}b^{2}}{(E-E_{1'})}.
\end{equation} 
\subsubsection{Crossing conditions}\label{cs}
   In general, the identification of the intermediate states of the $A+2n$ compound nucleus as IBM eigenstates - the unitary condition - gives also the crossing conditions of the open with closed channels. 
   
 In cold atoms, intermediate states are manifested experimentally through the aid of a magnetic field. The energy of the open channel is controlled by an external magnetic field which is used in order to trap the cold atoms. However, the atomic nucleus is bounded by the strong interactions and in $H_{c}$ the trap is represented by the IBM Hamiltonian over the collective states. 
 
 In the application of the IBM to the compound nucleus the role of the magnetic field is being played by the boson number of the reaction channel. This is implied by comparing the width obtained in intermediate molecular states of cold atoms \cite{Timmermans} with the width of the intermediate state in our case (\ref{e3400a}). That is 
 \begin{equation}\label{cc1}
 	\begin{split}
 	b=& \langle N_{b}+1| s^{\dagger}|N_{b}\rangle \times \int dr \Psi_{m}(r)u^{+}_{0}(r), \\
 	 & \langle S_{j}|V_{hf}|S_{in} \rangle \times \int d^{3}r \phi_{m}(r)u_{N}(r).
 	 \end{split}
 \end{equation}
 The first row shows the coupling of the width in the coupled channels equations of $H_{c}$ as presented in Eqs (\ref{e3400a}), (\ref{e3400b}). The second row shows the coupling of the width in systems of cold atoms as presented in \cite{Timmermans} where the intermediate state is a molecular state. In both cases the coupling part $b$ of the width $\Gamma$ depends on a matrix element over the channel states times the overlap between the open channel state with the intermediate state. In the experiment of colliding cold atoms, the open-closed channel crossing is achieved by the tuning of the hyperfine splitting $V_{hf}$ via an external magnetic field that couples the initial open channel spin state $S_{in}$ with the closed molecular channel spin state $S_{j}$. In the same sense, the matrix element of the first row corresponds to the tuning of the boson number of the target states which has been set exactly to the $s^{\dagger}$ that achieves the crossing with the first closed channel. Therefore the intermediate state is represented by the creation of one $s$ boson. However, the applied $F$ term in $H_{c}$ is $s^{\dagger}+s$ where the $s$ part annihilates the first closed channel target state giving back the open channel. This annihilation is in formal analogy with the second hyperfine spin-flip interaction which dissociates the intermediate molecule in cold atoms \cite{Timmermans}.
 
   In general, during a scattering of a stream of neutron pairs with an even-even heavy target, the tuning of the number of incident pairs in the open channel achieves the crossing energy with the $m$-th intermediate state. This is illustrated by the difference $\Delta$ between two states of the open channel. In our case that difference is measured with respect to the $N_{b}$ level. The correspondence of the magnetic field $B$ with the boson number is written down as
 \begin{equation}
 \Delta=E_{m}+\frac{\partial \Delta}{\partial B}(B-B_{m}) \rightarrow N_{b}+m +\frac{\partial \Delta}{\partial N'}(N'-m).
 \end{equation}
  Left of the arrow is the tuning $\Delta$ of the energy of the open channel in systems of cold atoms \cite{Timmermans}. When the value of the external magnetic field $B$ takes the critical value $B_{m}$, $\Delta$ succeeds the crossing energy $E_{m}$. Right of the arrow, the crossing energy is $E_{m}=N_{b}+m$ and the number of the external incident neutron pairs is denoted by $N'$. In the same sense, when the number of the external incident pairs $N'$ equals the critical value of the $m$-th intermediate state, the crossing energy of the open with the closed channel is achieved. The form of the two-neutron transfer operator $s^{\dagger}+s$ for the $H_{c}$ is the example of the tuning for one incident neutron pair. 
  
  While collision channels between cold atoms are classified by the overall magnetic spin of the state - and therefore are tuned with respect to the external magnetic field - the reaction channels in our case are classified with respect to the boson number. In general, the action of the magnetic field on the Feshbach resonance of cold atoms is an example of how a gauge field affects the crossing of the open with a closed channel. In the IBM, crossing conditions involve variations of the boson number. By introducing the one dimensional conformal group in the IBM below, the boson number of reaction channels resembles a generalised $SU(1,1)$ spin and this implication is discussed briefly in subsection \ref{3}.
\section{Conformal symmetry} \label{conformal}
When the neutrons  enter the vicinity of the collective state there is the probability of their capture and the formation of the compound nucleus which is represented by the intermediate state $|c_{m} \rangle$. In analogy with cold atoms, this is a trapped state in a trap provided by the IBM Hamiltonian $H(\rho)$. The trap of the hyperspherical Eq (\ref{WC}) has a radius which at the eve of the resonant coupling is roughly equal with the radius of the boson number of the IBM, i.e $R^{2} \sim \langle \rho^{2} \rangle$. The mean value of the IBM Hamiltonian $H(\rho)$ over the target states offers a trap of the form $(1/2)M \omega^{2} \langle \rho^{2} \rangle$. It is better to call it as an effective trap since $H(\rho)$ contains in addition the kinetic term of the low-lying collective excitations. The energy of the trapped states $E_{m}$ results from the mean value of $H(\rho)$. There are $n$ such "trapped" states in the whole closed channels space $Q$ which is spanned by the IBM states with $N_{b}+1, N_{b}+2$ bosons and so on.
 
 In what follows the generators of the $SO(2,1)$ group are defined for the IBM in subsection \ref{symm}. A tower of equally spaced states emerges and subsection \ref{mts} comments on the measurement of these states. Finally, theoretical remarks are addressed in subsection \ref{5}.
\subsection{$SO(2,1)$ group in the IBM}\label{symm}
The solutions of the IBM equation (\ref{Mex}) read
\begin{equation}\label{c1}
	\Phi_{n}(\rho)=\frac{F^{\sigma}_{J}(\rho)}{\rho^{5/2}}, \quad F^{\sigma}_{J}(\rho)=\frac{\rho^{\sigma}}{a_{ho}^{\sigma}} L^{\sigma+2}_{J}\left(\frac{\rho^{2}}{a^{2}_{ho}}\right) e^{-\rho^{2}/2a^{2}_{ho}},
\end{equation}
with  eigenvalues
\begin{equation}\label{c2}
	E(N_{b})=\left(\sigma+2 J +\frac{6}{2} \right) \hbar \omega.
\end{equation} 
The oscillator length is $a_{ho}=\sqrt{\hbar / M \omega}$ and $ L^{\sigma+2}_{J}(\rho^{2}/a^{2}_{ho})$ is the associated Laguerre polynomial. The number of bosons obeys the relation $N_{b}=\sigma+2J$ with $J$ to classify the representations for a specific value of $N_{b}$.

One writes down the energy scale of the reaction channels in terms of the difference $E_{n}-E_{0}$. That is $E_{1}-E_{0}=E(N_{b}+1)-E(N_{b})=\hbar\omega$ in the first closed channel, $2\hbar\omega$ in the second closed channel and so on. This energy difference is the energy cost to separate a pair of neutrons from a nucleus. Therefore the frequency of the trap is given by $\hbar \omega=S_{2n}$ the two neutron separation energy. 

On the other hand, the harmonic oscillator length $a_{ho}$ adjusts the IBM Hamiltonian to the neutron mass $M$ of the incident neutron pair. Namely, the scaling of the radial coordinate $\rho/a_{ho}$ and accordingly of the six cartesian coordinates $q_{i}/a_{ho}$ gives length units to the boson number radius $\rho$. The length units are determined by the neutron mass $M$ and the two neutron separation energy $\omega=S_{2n}/\hbar$.

It is useful now to introduce the generators of the one dimensional conformal group in the IBM. After this introduction relations that occur in the unitary limit of the cold atoms \cite{WC} are easily obtained in the IBM.
The bosonic expressions of $H$, $D$ and $K$ are revealed by the usual canonical transformation of the coordinates and momenta \cite{Mexico}. 
The cartesian forms are
\begin{equation}\label{c3}
	\begin{split}
		& H=\sum_{i=0}^{5}-\frac{\hbar^{2}}{2M}\partial^{2}_{i}, \quad K=\sum_{i=0}^{5}\frac{1}{2} M \omega^{2} q^{2}_{i}, \\ &  D= \sum_{j=0}^{5}\frac{1}{2i}\left(\partial_{j} q_{j}+q_{j} \partial_{j}\right)=\frac{6}{2i}-i\rho\partial_{\rho}.
	\end{split}
\end{equation}
These operators satisfy the commutation relations of Eq  (\ref{e6}).
The special conformal operator is 
\begin{equation}\label{c4}
	K=\frac{\hbar \omega}{4}\left(({\bf d}^{\dagger}+{\bf d})({\bf d}^{\dagger}+{\bf d})+(s^{\dagger}+s)(s^{\dagger}+s) \right), 
\end{equation}
the dilatation operator is
\begin{equation}\label{c5}
	\begin{split}
		&D=\frac{-i}{4}((s-s^{\dagger})(s+s^{\dagger})+ (s+s^{\dagger})(s-s^{\dagger})+ \\
		&(\bf{d}-\bf{d}^{\dagger})(\bf{d}+\bf{d}^{\dagger})+(\bf{d}^{\dagger}+\bf{d})(\bf{d}-\bf{d}^{\dagger})),
	\end{split}
\end{equation}
and the free space Hamiltonian is
\begin{equation}\label{c6}
	H= -\frac{\hbar \omega}{4}\left((s-s^{\dagger})(s-s^{\dagger})+(\bf{d}-\bf{d}^{\dagger})(\bf{d}-\bf{d}^{\dagger})\right).
\end{equation}
These expressions give 
\begin{equation}\label{c41}
K=\frac{\hbar \omega}{4}\left({\bf d}^{\dagger}{\bf d}^{\dagger}+{\bf d}{\bf d}+s^{\dagger}s^{\dagger}+ss+ 2{\bf d}^{\dagger}{\bf d}+2s^{\dagger}s+6 \right), 
\end{equation}
\begin{equation}\label{c51}
D=\frac{i}{2}\left(s^{\dagger}s^{\dagger}-ss+{\bf d}^{\dagger}{\bf d}^{\dagger}-{\bf d}{\bf d} \right), 
\end{equation}
\begin{equation}\label{c61}
H= \frac{\hbar \omega}{4}\left(-{\bf d}^{\dagger}{\bf d}^{\dagger}-{\bf d}{\bf d}-s^{\dagger}s^{\dagger}-ss+ 2{\bf d}^{\dagger}{\bf d}+2s^{\dagger}s+6 \right).
\end{equation}
One derives now the linear combinations
\begin{equation}\label{c42}
H+K=\hbar \omega\left({\bf d}^{\dagger}{\bf d}+s^{\dagger}s+ \frac{6}{2} \right), 
\end{equation}
\begin{equation}\label{c52}
H-K=-\frac{\hbar \omega}{2}\left({\bf d}^{\dagger}{\bf d}^{\dagger}+{\bf d}{\bf d}+s^{\dagger}s^{\dagger}+ ss \right). 
\end{equation}
The commutation relations of the $SO(2,1)$ group \cite{WC, Pitaevskii} are closed by three generators $L_{1}$, $L_{2}$, $L_{0}$ as they are defined by
\begin{equation}\label{c9}
2L_{1}=\frac{1}{\hbar \omega}(H-K), \quad 2L_{2}=D, \quad 2L_{0}=\frac{1}{\hbar \omega}(H+K).
\end{equation}
Ladder operators \cite{WC, Pitaevskii} are defined by the relation $L_{\pm}=2(L_{1} \pm iL_{2})$ which gives
\begin{equation}\label{c8}
L_{\pm} = \pm iD+ \frac{1}{ \hbar \omega}\left(H-K\right),
\end{equation}
and reads in terms of bosons
\begin{equation}\label{c81}
L_{+} = -({\bf d}^{\dagger}{\bf d}^{\dagger}+ s^{\dagger}s^{\dagger}), \quad L_{-}=-({\bf d}{\bf d}+ss).
\end{equation}
Having defined these generators the IBM Hamiltonian $H(\rho)$ is provided by the combination $H+K$ which corresponds to the $L_{0}$ operator. The dilatation operator $D$ and the combination $H-K$ correspond to the non-compact rotations or boosts generated by $L_{2}$ and $L_{1}$ respectively. These operators as well as the ladder $L_{\pm}$ create or annihilate states with {\it two} bosons on the eigenstates of the IBM equation.

 One focuses now on the eigenstates of the IBM equation in Eqs (\ref{c1}) and (\ref{c2}). Each eigenstate has a specific boson number and represents a stationary state of frequency $\omega$. A formal solution is defined for $\omega=0$ and $E(N_{b})=0$. This is a limit of the spectrum (\ref{c2}) in which the lower bound of $(6/2)\hbar \omega$ from the $U(6)$ is absent and is amenable to be interpreted as a resonance state. 
In relevance to a reaction channel which constitutes of the incident (outgoing) two neutrons times the target state, one examines now the IBM state of the first closed channel with $N_{b}+1$ bosons.
 A zero energy or free space eigenstate ($\omega=0$) that corresponds to the $N_{b}+1$ boson number is the resonance state describing the crossing of the first closed channel with the open channel of two incident neutrons onto the $N_{b}$ level.

  In the terminology of cold atoms \cite{WC}, states within the trap are mapped to free space eigenstates denoted as $\psi^{0}_{\nu}$ and vice-versa. Such a mapping is realized now for the Hamiltonian $H_{c}$ of the $A+2n$ compound nucleus where the trapped (intermediate) states are provided by the states of the IBM equation in the unitary limit. One of the results of the algebraic comparison of section \ref{2} is that the scaling exponent $\nu$ of the mapping coincides with the $O(6)$ quantum number $\lambda$ for the zero energy state $\psi^{0}_{\nu}$. Because of the algebraic correspondence $\lambda \rightarrow \sigma$ explained in section \ref{2}, the scaling exponent of the free space eigenstate of the IBM equation is $\nu=\sigma$. That state is written as $\psi^{0}_{\sigma}$ and corresponds to a specific boson number with $\sigma=N_{b}-2J$ in general. 
  
  The channel state $|c_{1} \rangle$ consists of the amplitude $\Psi_{1}(r)$ and the target state $|\Phi_{1}(\rho)\rangle$ of specific label $\sigma,J$. The target state is obtained from the free space eigenstate $\psi^{0}_{\sigma}$ through the mapping
\begin{equation}\label{c7}
	|\Phi_{1}(\rho) \rangle = (L_{+})^{J}e^{-\rho^{2}/2 a^{2}_{ho}} |\psi^{0}_{\sigma} \rangle.
\end{equation}
 The exponential function $exp(-\rho^{2}/2a^{2}_{ho})=exp(-\hbar \omega K)$ bounds the free space eigenstate and contains the special conformal operator $K$. 
 The form of the free space eigenstate $\psi^{0}_{\sigma}(\rho)$ is obtained by acting with $L_{-}$ on the lowest energy eigenstate. For the first closed channel, the lowest energy eigenstate has $ \sigma=N_{b}+1$ in the representation with $J=0$. One performs the same steps as in \cite{WC} and the free space wavefunction is
\begin{equation}\label{c10}
	\psi^{0}_{\sigma}(\rho, \Omega_{5})=\rho^{N_{b}+1} Y(\Omega_{5}).
\end{equation} 
If one applies this wavefunction back to the IBM equation (\ref{Mex}) the result for $\omega=0$ is a zero energy state for $N_{b}+1$ bosons. Therefore,  the analog of the scaling relation of Eq (\ref{e4}) is Eq (\ref{c10}) which defines the scaling exponent of the zero energy state $\psi^{0}_{\sigma}$ for the IBM. In other words, the scaling exponent of a nuclear collective state is determined by the boson number in the IBM.

 Under a scaling of the boson number radius of the form $\tilde{\rho}=\rho/\lambda$, the free space eigenstate behaves as
\begin{equation}\label{c10a}
\psi^{0}_{\sigma}(\tilde{\rho},\Omega_{5})=\frac{1}{\lambda^{N_{b}+1}}\psi^{0}_{\sigma}(\rho, \Omega_{5}).
\end{equation}
As presented in Appendix \ref{appBC}, a scaling of the target state does not affect the boundary condition for the pair wavefunction. This situation defines the boundary condition of the open channel wavefunction. Similarly, a boundary condition is derived for the closed channel wavefunction through the mapping from the free space eigenstate. Under a scaling transformation of the target state $\Phi_{1}({\tilde\rho})$, the boundary condition for the first closed channel wavefunction behaves as
\begin{equation}\label{c10b}
\lim_{r \rightarrow 0}\Psi_{1}(\tilde{\rho},r)=e^{-\tilde{\rho}/2a^{2}_{ho}}\psi^{0}_{\sigma}(\tilde{\rho},\Omega_{5})\frac{1}{r^{4}},
\end{equation}
with the pair wavefunction to be $\sim 1/r^{4}$ for $r\rightarrow 0$ and at the resonance where $a_{r} \rightarrow \infty$. The boundary condition for the closed channel wavefunction $\Psi_{1}(\rho,r)$ is provided by Eq (\ref{c10b}) through the use of the free space eigenstate $\psi^{0}_{\sigma}(\rho,\Omega_{5})$ and consequently a boundary condition of the same type applies to every closed channel wavefunction $\Psi_{m}(\rho,r)$. The boundary condition keeps its form under the scaling independently of the magnitude of $\lambda$. This is a manifestation of scale invariance with the scaling exponent $\sigma=N_{b}+1$ as the eigenvalue of the dilatation operator $D$ for the first closed channel.

However, the mapping of the free space eigenstate $\psi^{0}_{\sigma}$ to the intermediate state of the compound nucleus is described by the generators of the $SO(2,1)$ group. This mapping is characterized by a higher invariance with respect to the scale invariance of the dilatation operator alone. That is conformal invariance in one dimension - time - in the sense that a new sequence of time-dependent trapped states result from the mapping and are grouped with respect to the scaling exponent $\sigma$ of the free space eigenstate. These states represent the $SO(2,1)$ group in the $A+2n$ compound nucleus and are presented in the next subsection.
\begin{figure}
	\includegraphics[scale=0.5]{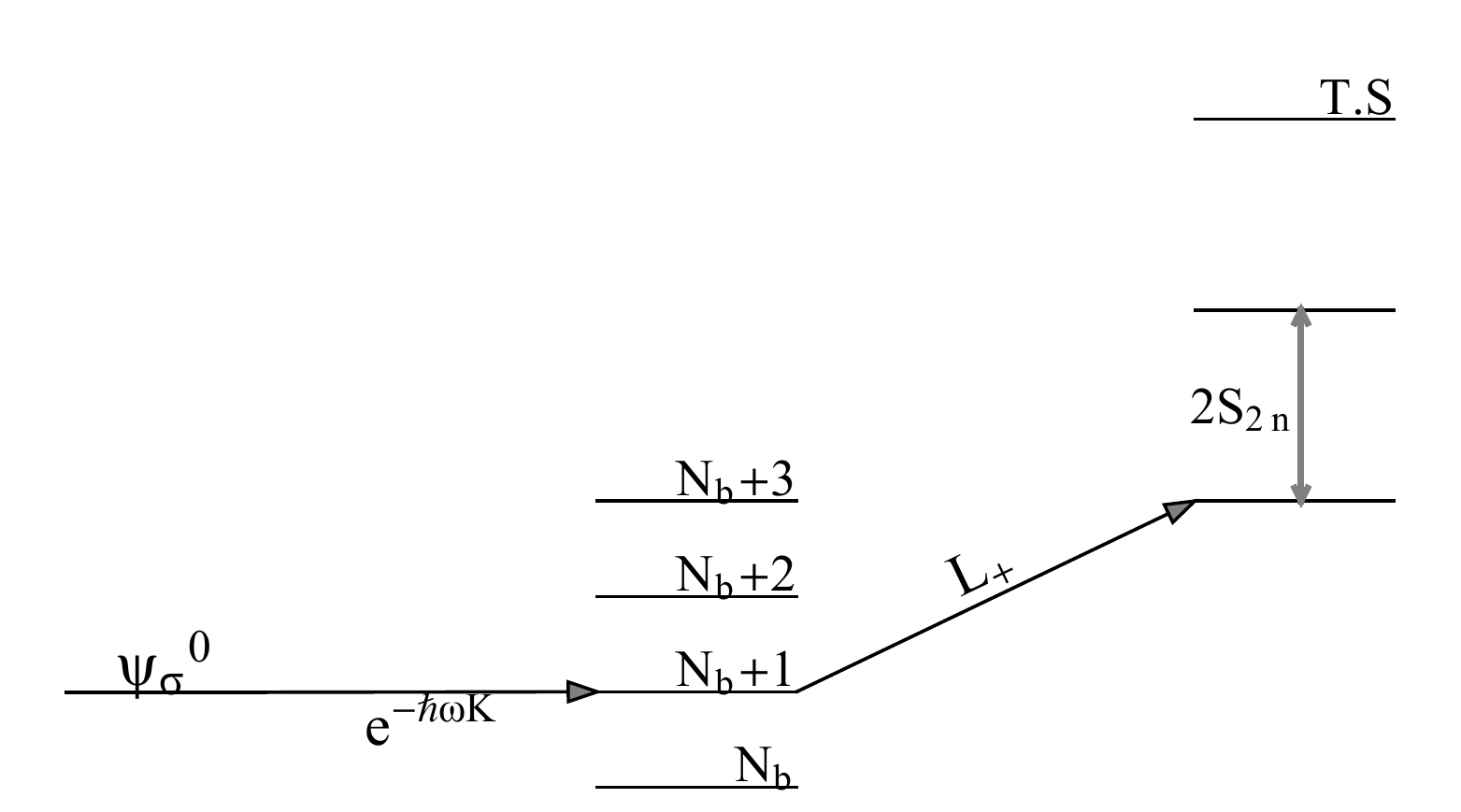}
	\caption{A graphical illustration of the action of the $SO(2,1)$ generators on the eigenstates of the IBM equation. In the scattering of $1$ pair of neutrons on a target nucleus, the zero energy state $\psi^{0}_{\sigma}$ appears as a resonance in the open channel. The scaling exponent $\sigma=N_{b}+1$ of the resonance state reflects the boson number of the closed channel eigenstate. By acting with $e^{-\hbar \omega K}$ on the resonance state $\psi^{0}_{\sigma}$ (left) of the first closed channel, the eigenstate of $N_{b}+1$ bosons is obtained (middle). By acting with $L^{k}_{+}$ on the $N_{b}+1$ eigenstate itself a tower of equally spaced states (right: T.S) appears with a spacing of $2$ times the two neutron separation energy $S_{2n}$. The eigenstate with $N_{b}+1$ bosons is the ground member of the tower.}
	\label{f1}
\end{figure}
\subsubsection{Tower of equally spaced states}\label{teq}
Of special interest in nuclear physics is the action of the $L_{+}$ generator $k$ times on the eigenstates of the IBM equation themselves. This occurs because of the commutation relation $[H+K,L_{+}]=2 \hbar \omega L_{+}$. By the repeated action of $L_{+}$ the spectrum of the IBM equation is coupled with a tower of equally spaced states of energy $2k \hbar \omega$. 

In the reaction channel that means the vector of each channel $|\Phi_{n}(\rho) \rangle$ is coupled to the whole tower of states
\begin{equation}\label{c11}
	(H+K)L^{k}_{+}|\Phi_{n}(\rho) \rangle=\left(E_{n}+2k \hbar \omega \right) L^{k}_{+}|\Phi_{n}(\rho) \rangle.
\end{equation} 
For instance, under the action of the $L_{+}$ generator $k$ times on the vector of the first closed channel $|\Phi_{1}(\rho) \rangle$ the result is the vector $L^{k}_{+}|\Phi_{1}(\rho) \rangle $. Therefore, the action of the generator $L^{k}_{+}$ on the channel state $|c_{1} \rangle$ gives
\begin{equation}\label{c12}
 L^{k}_{+} |c_{1} \rangle = \Psi_{1}(r)L^{k}_{+}|\Phi_{1}(\rho) \rangle.
\end{equation} 
One solution for the amplitude of the pair of neutrons $\Psi_{1}(r)$ in the first closed channel corresponds to $L_{+}|\Phi_{1}(\rho) \rangle$ as well as to $L^{2}_{+}|\Phi_{1}(\rho) \rangle$ and consequently to the $k$-th member of the tower. Namely, the intermediate state is coupled with a tower of equally spaced states.
 A target state of the tower is generated by the expression 
 \begin{equation}\label{ts}
 \Phi_{k}(\rho)=L^{k}_{+}\Phi_{1}(\rho).
 \end{equation} 
 The vector $|\Phi_{1}(\rho) \rangle$ defines the ground member of the tower. In the coupled channels equations the eigenvalues of the target states are
\begin{equation}\label{c13}
	\langle \Phi_{1}(\rho)|L^{k}_{-} H(\rho)L^{k}_{+}|\Phi_{1}(\rho) \rangle=E_{1}+2k\hbar\omega=S_{2n}+2kS_{2n}.
\end{equation}
A graphical illustration of the action of the $SO(2,1)$ generators is presented in Figure \ref{f1}. 

In the case of cold atoms confined in a trap, the action of $SO(2,1)$ generators creates similarly such a tower of equally spaced states \cite{WC}. The states of the tower result from a perturbation of the trap that is succeeded via a time-dependent scale factor in the classical wave limit of a boson gas \cite{Pitaevskii}. These time-dependent states result from the action of the time-dependent generators $L_{\pm}(t)=e^{\pm 2i\omega t}L_{\pm}$.

In the IBM, consider a time dependent target state $\Phi_{1}(\rho,t)$. The one dimensional conformal transformation is reflected on a local scale factor that depends only on time $\lambda(t)$. The time dependent scale factor $\lambda(t)$ signals the incident wave of neutrons that perturbs the boson number radius $\rho$. The transformation is the following \cite{WC,Pitaevskii}
\begin{equation}
	\tau(t)=\int_{0}^{t}\frac{dt'}{\lambda^{2}(t')}, \quad \tilde{\rho}=\frac{\rho}{\lambda(t)}, \quad \lambda(t)=\sqrt{1+\frac{E^{2}}{\hbar^{2}}t^{2}}.
\end{equation}
In the last equation, the form of the local scale factor $\lambda(t)$ is motivated by that of the cold atoms. In the vicinity of the resonance for the first closed channel $E=\hbar\omega=S_{2n}$. Like in cold atoms, the scaling occurs for $t<0$ and here reflects the preparation of the stationary state for scattering. 

In terms of the one dimensional conformal group \cite{Hagen}, the time dependence of a target state at the resonance is governed by the operator $G=uH'+vD+wK'$ with 
\begin{equation}
G|\Phi_{1}(\rho,t)\rangle=i\hbar f(t)|\dot{\Phi}_{1}(\rho,t)\rangle, \quad f(t)=u+vt+wt^{2}.
\end{equation} 
The primes in $H'=(M/\hbar^{2})H$ and $K'=K/(M\omega^{2})$ scale out the mass and frequency from the generators. In our case $u=1$, $v=0$ and $w=E^{2}/\hbar^{2}$ with $\lambda^{2}(t)=f(t)$.

 Writing the channel wavefunction in the form $\Phi_{1}(\tilde{\rho},\tau)$, the time dependent state reads
\begin{equation} \label{phi}
\Phi(\rho,t)= \frac{e^{\left(i M \rho^{2} \frac{\dot{\lambda}(t)}{2 \hbar \lambda(t)}\right)}}{\lambda(t)^{6/2}}\Phi_{1}(\tilde{\rho},\tau).
\end{equation} 
Tower states (\ref{ts}) are obtained from the expansion of this wavefunction in terms of infinitesimal changes in the local scale factor $\delta \lambda(t)$ \cite{WC}. That is
\begin{equation} \label{phi1}
\begin{split}
\Phi(\rho,t)= (e^{-iEt/\hbar}-\epsilon e^{-i(E+2\hbar \omega)t/\hbar}L_{+} \\  +\epsilon^{*}e^{-i(E-2\hbar \omega)t/\hbar}L_{-} )\Phi_{1}(\rho,0).
\end{split}
\end{equation} 
The time dependent states of the tower are therefore obtained from the action of the ladder operators that create or annihilate states with two bosons in the IBM. These states span the representations of the $SO(2,1)$ group \cite{Pitaevskii} that is isomorphic to the conformal group in one dimension.  

Experimentally, members of the tower of equally spaced states occur in  Bose-Einstein Condensates of cold atoms in elongated traps that are perturbed along a transverse direction via a transverse magnetic field \cite{Chevy}. The condensate exhibits two types of modes: A longitudinal along the elongation of frequency $\omega_{\parallel}$ and a transverse mode of frequency $\omega_{\perp}$. The frequency of the first member of the tower is $2\omega_{\perp}$ and occurs as a result of the pertubation. 

Based on the picture of this experiment from trapped cold atoms, the realization of the $SO(2,1)$ mapping of the free space eigenstate to intermediate states in nuclear physics is completed by the estimation of the small parameter $\epsilon$ of the expansion (\ref{phi1}). This occurs below for the tentative application of the representations of the $SO(2,1)$ group the IBM-compound Hamiltonian $H_{c}$ through the two-neutron transfer.

 In the target states, the analog of two types of modes of the boson condensate is indicated by the eigenvalues of the full IBM Hamiltonian that includes two-body terms in the $O(6)$ limit \cite{IBM, Mexico}
\begin{equation} \label{mv}
	E_{n}=\epsilon_{0}-A\frac{1}{4}\sigma(\sigma+4)+B \frac{1}{6}\tau(\tau+3)+ C L(L+1).
\end{equation} 
$\epsilon_{0}$ is at the order of the binding energy of the target nucleus as a whole. A change in the binding energy $\epsilon_{0}$ caused by the addition of one boson equals to $S_{2n}$ and plays the role of $\omega_{\perp}$. The rest terms reflect vibrational and rotational low-lying collective excitations caused by the quadrupole deformation. A typical frequency of these low-lying collective excitations is the analog of the elongated mode $\omega_{\parallel}$.  

For example, take as $\omega_{\parallel}$ the $E(2^{+}_{1})$ of the target nucleus. Like in cold atoms, infinitesimal changes in the scale factor $\lambda(t)=1+\delta\lambda(t)$, $\delta\lambda(t)<<1$, cause an oscillation of the boson number radius $\rho(t)=(1+\delta\lambda(t))\rho(0)$. By choosing the small parameter $\epsilon=E(2^{+}_{1})/S_{2n}$, the boson radius oscillates as \cite{WC} 
\begin{equation}
	\delta\lambda(t)=\epsilon e^{-2i(S_{2n}/\hbar) t}+\epsilon e^{2i(S_{2n}/\hbar) t}+O(\epsilon^{2}).
\end{equation}
This is an example of how the mode of energy $E_{n}\pm2S_{2n}$ appears in the energy spectrum of the compound nucleus 
in the sense of a classical wave limit of a boson gas \cite{WC,Pitaevskii} that is
perturbed by two incident (emergent) neutrons. 
\subsection{Measurement of tower states}\label{mts}
So far the width and the position of the energy of the intermediate state has been determined. In the two-neutron transfer to the ground state with $N_{b}$ bosons, the intermediate state appears as a fluctuation to the cross section as seen from relation (\ref{20}). The relevant cross section that manifests this fluctuation is the compound-elastic
\begin{equation}
\sigma_{ce}=\frac{(4\pi)^{3}}{k^{2}_{r}} \frac{\Gamma^{2}_{1}}{(E-E_{1})^{2}+(\Gamma_{1})^{2}/4},
\end{equation}
 and is derived in the appendix \ref{appcs}. This cross section is experimentally derivable from the difference between the total and the direct cross section. $E_{1}$ refers to the energy of the first closed channel and accordingly the width is
 \begin{equation}\label{e3401a}
	\Gamma_{1}=b^{2}_{1} \left(\frac{4 M} {\hbar^{2}}\right) k_{r},
\end{equation}
  with $b^{2}_{1}=(N_{b}+1) \abs{ \int dr \Psi_{1} u^{+}_{0}(r)}^{2}$. The width is normalized to the energy scale of $S_{2n}$. However, before comparison with the experiment one needs to average the scattering matrix, transition amplitude and cross sections. This is a typical procedure for measuring fluctuations of cross sections in compound nuclei. The computation of the average involves the upper and the lower limit of the integration. Of interest is the lower limit as these fluctuations reveal a fine structure of the total cross section in the low energy limit of the integration interval. In the pair-collective state resonance, the average interval $\Delta E$ should be larger than the width of the intermediate state i.e $\Delta E > \Gamma_{1}$.
 
   The upper limit of the integration average is to be determined from the experiment and constitutes that value which is much larger than $\Gamma_{1}$ but does not change significantly the average cross section. The maximum value of the $\sigma_{ce}$ reflects the maximum value of the effective scattering length in Eq (\ref{21}) and indicates the unitary limit in the pair-collective state resonance in conjuction with the lowest value of $\Gamma_{1}$. The value of the width is also determined by the relative kinetic energy $k_{r}$ which for very low values zero gives a rather narrow fluctuation. 

The comments above apply to the ground member of the tower which corresponds to the first closed channel. For the measurement of the excited states of the tower one argues that the first member is an excited state by two bosons more with respect to the ground member of the tower with $N_{b}+1$ bosons. Therefore this state may be perceived as an intermediate state with $N_{b}+3$ bosons. In order to measure this state one chooses the zero energy level as a ground state with $N_{b}+2$ bosons. In this case the open channel is that with $N_{b}+2$ bosons and its coupling with the incident pair of neutrons creates a resonance as the $N_{b}+3$ level of the IBM. Then the observation of an intermediate state at energy $(N_{b}+3) S_{2n}$ of width $\Gamma_{3}$ where now $b^{2}_{3}=(N_{b}+3) \int dr \Psi_{2} u^{+}_{0}(r)$ indicates the first excited member of the tower emerging from the target with $N_{b}$ bosons. This resonance should appear as a fluctuation to the cross section  of the two neutron transfer to the isotope with a ground state of $N_{b}+2$ bosons. Accordingly, the resolution of this fluctuation should be performed in the energy interval with a lower limit that is larger than $\Delta E> \Gamma_{3}$ and the upper limit  is to be determined by the experiment as before. However, due to arguments presented in subsection \ref{3}, in a large $N_{b}$ limit one expects that the width will remain constant for each member of the tower. In such a case $\Gamma_{3}=\Gamma_{1}=\Gamma_{n}$, which implies a constant value of the energy interval as $\Delta E>\Gamma_{n}$ and the pattern of fluctuations becomes even more regular with respect to their widths. 
 
 In a series of isotopes one therefore expects to see a regularity pattern in the fluctuations of cross sections with respect to the mass number of the target nucleus. If the two-neutron separation energy remains approximately constant along these isotopes, conformal symmetry is represented on a regularity pattern of the cross section fluctuations - both in their positions and widths with respect to the mass number.
 
 The degeneracies of the fluctuations emerging from the isotope with a ground state of $N_{b}+2$ bosons with the tower states emerging from the isotope with a ground state of $N_{b}$ bosons may be tested by varying the number of the incident neutron pairs with respect to the initial target nucleus with the ground state of $N_{b}$ bosons. In the limit of a very low kinetic energy of the transferred pairs, such a number variation is equivalent with the variation of the total energy of the channel. Namely, the position and the width of a resonance formed by 3 incident neutron pairs of total energy $3S_{2n}$ on the isotope with $N_{b}$ bosons should be degenerate with the energy of the intermediate state emerging from the scattering of one neutron pair of total energy $S_{2n}$ on the isotope with the ground state of $N_{b}+2$ bosons. These degeneracies result from the representations of the one dimensional conformal group in the IBM-compound Hamiltonian $H_{c}$ for the two-neutron transfer paradigm. In such a case the regularity pattern of the fluctuations of the cross section appears with respect to the number of ingoing (outgoing) neutrons.
\subsection{Theoretical remarks}\label{5}
This subsection is structured into three parts. The first part comments briefly on the introduced  coupling $F$ term and on the unitary limit of the IBM-compound Hamiltonian in relation with bound states in the continuum \cite{Weidenmuller, Lipperheide}. It also briefly comments on the unitary pair-collective state interaction and its associated scattering length in relevance to a unitary boson-boson interaction. The second part remarks on some theoretical consequences that follow the introduction of the unitary limit in heavy nuclei. These start from the relation of an underlying critical point with the fluctuations of the cross section on the one hand and with the conformal algebra on the other. They also include the manifestation of the BCS-BEC crossover in compound nuclei. The third part contains some general comments on the initial question about the relation of the symmetries of the IBM with the symmetries of gauge groups.
 \subsubsection{First part}
The unitary limit of the IBM-compound Hamiltonian $H_{c}$ is now compared with the concepts of bound states in the continuum and doorway states \cite{Weidenmuller,Lipperheide}. In this paper no assumption for a particular type of state or ansatz of states, such as the doorway states, was performed in order to interpret the intermediate states of the $A+2n$ compound system. Intermediate states were applied in their most general sense - states that manifest themselves between the entrance and the exit channel as fluctuations of the average of cross section  - in the IBM-compound $H_{c}$ restricting them however to the unitary limit. The latter was revealed by the (unitary) condition that the energy of the intermediate state coincides with the energy of the IBM state of the closed channel. In the calculations of the Appendix \ref{app11}, the continuum of the open channel is manifested on the expansion of the propagator $g^{+}_{p}(E)$ over the open channel states.

 In the general case - away from the unitary condition - the continuum is present in the scattering state $\Psi_{0}(r)$ of the open channel set $P$ as well as in the scattering states $\Psi_{m}(r)$ of the closed channels set $Q$. Both of these sets contain the scattering wavefunction and the wavevector $k_{r}$ may be expressed as $k_{r0}$ for the open and and $k_{rm}$ for the closed channels. Then the operators $P$ and $Q$ are written as densities in the continuum
\begin{equation}  \label{t}
\begin{split}
& P=|\Phi_{0}(\rho)\rangle \langle \Phi_{0}(\rho)| \int dk_{r0} |k_{r0}\rangle\langle k_{r0}|, \\
& Q=|\Phi_{m}(\rho)\rangle \langle \Phi_{m}(\rho)| \int dk_{rm} |k_{rm}\rangle\langle k_{rm}|.
\end{split}
\end{equation}
 In such a case, the unitary condition is equivalent with the following two steps. First, in addition to the sets of Eq (\ref{t}), one introduces a third set of pure IBM states for the closed channels space with energies determined by the constancy of the $S_{2n}$. Second, one chooses the average integral to be performed around the determined energies of the third set of pure IBM states for the entrance and the exit channel. The closed channels set separates as $Q=t+q$ where $t=|\Phi_{t}\rangle \langle \Phi_{t}|$ consists of the third set of pure IBM states - the tower states - without being multiplied to the scattering states of the pair. The remaining part of the closed channels set is now the density $q= |\Phi_{m}(\rho)\rangle \langle \Phi_{m}(\rho)| \int dk_{rm} |k_{rm}\rangle\langle k_{rm}|$ with the wavevector $k_{rm}$ to be in general arbitrary. In such a framework, the unitary condition is equivalent to the zero values of $k_{r0}$ and $k_{rm}$ which determines the energy of the continuum states to be those of the IBM states  $t$. Acordingly, the channel couplings - the matrix elements of the $F$ term in the continuum - carry an energy density that is normalized to the two-neutron separation energy $S_{2n}$. Along with the channel couplings, $t$ states determine the energy range of the fluctuations of the corss section i.e the average integral. By this equivalence one infers that the states involved in the unitary limit of the pair-collective state scattering behave as pure IBM states in the continuum. 
 
 The third set of pure IBM states is to be compared with relevant studies of one neutron resonances in the continuum Shell Model where the third set of states is named as bound states in the continuum \cite{Lipperheide, Weidenmuller}. There, the channel couplings - the analog of the matrix elements of the $F$ term - are provided by the residual interaction. Here the bound states in the continuum emerge from the unitary condition which interprets the intermediate states as boson states i.e pure IBM states. This interpretation takes place  in analogy with the coupled channels in systems of fermionic cold atoms where the intermediate states are manifested as bosonic-molecular states. 

The use of the doorway state formalism was avoided in this paper due to particular assumptions in the couplings $H_{tP}$ and $H_{Pq}$ which are compensated here by the more general $H_{PQ}$ coupling that applies to the compound-elastic reaction and incorporates the unitary limit. The determined positions of the intermediate states in the unitary limit are therefore simply attributed to the IBM symmetries, to the constancy of the two neutron separation energy and to the $SO(2,1)$ group.

For the pair-collective state interaction it deserves to be mentioned that it may be reduced to a unitary boson-boson interaction in the IBM. If the collective state is formed by one boson, the delta function $\delta(\rho-R)$ refers to the interaction of the incident pair localised at $R$ with one valence boson localised at $\rho$. When the pair is trapped it behaves like one $s$ boson and in that case the scattering length $a_{r}$ refers to a boson-boson interaction. In analogy with cold atoms, the boson-boson interaction is the molecule-molecule interaction of trapped cold fermionic atoms \cite{Pricoupenko}. 

However, the neutron-neutron scattering length $a$ controls the strength of the delta interaction $\delta(r_{1}-r_{2})$ of the incident neutrons themselves. In the unitary limit of cold atoms, the scattering length of the molecule-molecule (boson-boson) coupling is reduced to the binary atom-atom (nucleon-nucleon) scattering lengths \cite{Pricoupenko}. Since the interaction strength is maximum for all the particles involved in the unitary limit, the neutron-neutron scattering length $a$ should maximize along with the maiximization of $a_{r}$. The pair-collective state resonance reflects the mapping of $N=2$ unitary neutrons from the open channel to the state of the trap consisting of $N_{b}+1$ bosons in the closed channel. 

The delta function $\delta(\rho-R)$ is one dimensional in the $O(6)$ limit. It is interesting to write the form of this function in the nucleonic language. In such a case this interaction approximates the unitary interaction between each incident neutron with each target nucleon of the valence pairs (bosons) that form the collective state. In the case where the collective state is represented by one valence boson, the interaction is written approximately as $\delta(r_{1}-r_{1}')\delta(r_{2}-r_{2}')$ with the prime coordinates to denote the target nucleons of the valence boson. This is a six-dimensional delta function \cite{Morse}. In this form one writes the magnitude of the boson number radius as $\rho^{2}=r'^{2}_{1}+r'^{2}_{2}=q^{2}_{0}+\beta^{2}$, an expression which does not involve the angles. 
\subsubsection{Second part}
One now starts a brief discussion for the consequences that follow the introduction of the unitary limit in heavy nuclei. Some of them are discussed here related with the candidacy of the $A+2n$ compound nucleus  at the unitary limit for a critical point and for the manifestation of the BCS-BEC crossover.

The unitary condition produces the maximum value of the fluctuations around the average value of the cross section. Such a situation introduces the $A+2n$ compound nucleus in the unitary limit as a candidate for the manifestation of an underlying critical point. In other words one implies a relation between the unitary condition and a critical point that is experimentally manifested via the maximization of the fluctuations of the cross section. 

 In general the symmetry of a critical point is expressed via the conformal algebra which here produces not only the maximization of the fluctuations, but also their regularity pattern with respect to the mass number of the target nucleus. The fluctuations of the cross section are determined by the matrix elements over the $F=s^{\dagger}+s$ term. That $F$ term commutes with the special conformal operator $K$ and creates the zero energy state $\psi^{0}_{\sigma}$. In other words, that $F$ term behaves as a primary operator of the conformal algebra that is isomorphic to the $SO(2,1)$ group. One then obtains the energy of the intermediate state in the unitary limit by the scaling dimension of the primary operator. This is the so called operator-state correspondence \cite{Son}. In such a case, the role of the primary state is being played by the free space eigenstate $\psi^{0}_{\sigma}$ and the $s^{\dagger}+s$ - the coupling that achieves the crossing of the open with the closed channel - is the primary operator that creates this state. The scaling exponent $\sigma$ gives the boson number of the IBM state at resonance which upon trapping (capture) gives the energy of the trapped (intermediate) state of the compound nucleus.

On the other hand, the $s$ boson is a pair of two nucleons and this resembles the dimer creation operator - the coupling of two unitary nucleons - which has scaling dimension $\Delta=2$. For the case of two incident nucleons, the dimer operator is of the form $\lim_{r_{1}\rightarrow r_{2}}\psi(r_{1})\psi(r_{2})/|r_{1}-r_{2}|$ \cite{Son} as a scalar coupling to total angular momentum zero. In a non-relativistic Conformal Field Theory, these creation operators act on the vacuum to create a scale invariant state such as the $\psi^{0}_{\sigma}$ in our case. Each creation operator corresponds to a state of energy $\Delta \hbar \omega$ that is created by multiplying with the exponential $e^{-H}\psi^{0}_{\sigma}$. The conformal algebra is realized through the $L_{\pm}$ operators and creates a tower of such equally spaced states. The observation of a regularity pattern in the fluctuations of the elastic cross section along a series of isotopes should exemplify the operator-state correspondence for the dimer of two unitary neutrons with $\Delta=2$. This mapping refers to the time-dependent trap and here produces the energy of the excited members of the tower. The energy of the ground member of the tower is provided by the scaling exponent $\sigma$.

The unitary limit draws the attention to the manifestation of the BCS-BEC crossover \cite{Randeria} in the $A+2n$ compound nucleus. The crossover is manifested on the intermediate state that satisfies the unitary condition while the BCS and the BEC limits refer to intermediate states away from the unitary condition. The quantity of interest is $\xi=1/(k_{F}a) \sim 1/(\sqrt{N_{b}}a_{r})
$, where the Fermi momentum corresponds to the energy of the ground state of the target nucleus and is proportional to the square root of the boson number. In general this quantity is small in nuclear matter and one expects to be tuned by the compound elastic reaction reaching the crossover value of zero in the unitary limit. It is not now clear what physical situation corresponds to the BCS limit where $\xi \rightarrow - \infty$ and to the BEC limit where $\xi \rightarrow \infty$ in relevance to the $A+2n$ compound system. Both of these limits are away of the unitary limit.

It deserves to be mentioned that the IBM equation of the $O(6)$ limit should be compared with a Gross-Pitaevskii equation upon the addition of a unitary boson-boson interaction in the IBM.
 In molecules, the experimental achievement of a Bose Einstein Condensation is realized through the evaporative cooling of molecules in a trap \cite{Randeria}.  There, the atoms at resonance form diatomic molecules which eventually undergo a transition to a condensed molecular state.  In the $A+2n$ compound nucleus, the analog of a condensed state is an IBM state in the classical limit which eventually may be relevant to the evaporation decay of pairs of neutrons. The evaporation decay is beyond the scope of the present paper as it involves
  inelastic reactions with exit channels that differ from the entrance one.

\subsubsection{Third part} \label{3}

 We turn now to the initial question concerning the relation of the symmetries of collective nuclear states with the symmetries of gauge groups. A first approach to this relation is through field theoretical methods. In that case one examines a CFT where the unitary fermions are nucleons with an $SU(2)$ gauge group \cite{Son}. Such an approach is formally similar with the theory of Mukerjee and Nambu \cite{Nambu} for the IBM in which the transverse excitations of the $SU(2)$ gauge group refer to the $\sigma$ meson while the longitudinal ones refer to  the $\pi$ meson. Mukerjee and Nambu start from infinite nuclear matter evaluating bubble diagrams between the nucleons with an order of approximation provided by the cutoff of the renormalization group. The same round of calculations is performed in the relevant CFT \cite{Son} apart from the conformal algebra related with the unitary limit. One of the results of the present paper is the trapping of the $N=2$ unitary neutrons by the special conformal operator $K$ in the energy scale determined by the two neutron separation energy. This aspect of conformality bounds the energy from below - a limit that resembles the infrared fixed point of the renormalization group.

 However, the algebraic framework of the IBM permits to search for conformal algebras with infinite number of generators at the critical point that corresponds to the unitary limit. An example of such algebras emerges in a scattering case where the reaction channels are put into a 1-1 correspondence with the points of a unit circle \cite{Ramond}. In our case that means the realization of the $SO(2,1)$ generators on a unit circle, a process which gives rise to the representations of the $SU(1,1)$ group.
 Each point of this circle corresponds to a reaction channel state of specific boson number $N_{b}$ and pair momentum $k_{r}$. The energy of the channel state $|c_{n}\rangle=|k_{r};\lambda\rangle \otimes |N_{b},\sigma,\tau;L\rangle$ depends principally on the pair momentum $k_{r}$ which is continuous and on the boson number $N_{b}$ which is discrete. Their combination resembles the continuous and discrete representations of the $SU(1,1)$ group \cite{AGI}. In such a case the boson number resembles an $SU(1,1)$ spin in the discrete representation while $k_{r}$ corresponds to the $SU(1,1)$ spin of the continuous representation. In that sense the boson number and the pair momentum label the representations of a generalised $SU(1,1)$ spin on the unit circle.
 
 With respect to the fluctuations of the cross sections, the representations on the unit circle may be useful to impose further constraints on the widths of the fluctuations for the excited members of the tower. A permutation of the channels around the circle, means that the examined process $2n+N_{b} \rightarrow N_{b}+1 \rightarrow N_{b}+2n $ is compared with the process $2n+N'_{b} \rightarrow N'_{b}+1 \rightarrow N'_{b}+2n$, where $N'_{b}=N_{b}\pm m$. For instance, such a permutation applies to the measurement of the first excited member of the tower when the incident pair of neutrons scatters from the isotope with ground state $N_{b} \rightarrow N'_{b}=N_{b}+2$. Does the transition amplitude remains the same for these two processes? If yes, the width of the fluctuations remains constant along all the members of the tower. In the large $N_{b}$ limit of the IBM where $N'_{b} \sim N_{b}$ the transition amplitude approximately remains the same. 

 The way that this unit circle is mapped on to the $U(6)$ algebra of the IBM extends its Lie algebraic framework to algebras with infinite number of generators the so called Kac-Moody algebras  \cite{Franco}. In such a case one implies that algebras with infinite numbers of generators in the IBM may exist in a common algebraic framework with the large $N$ limit of the $SU(N)$ gauge group. This situation is implied by the impact of the boson number on the crossing condition which acts like the magnetic (gauge) field acts on the Feshbach resonance of cold atoms. That is, the boson number acts like a gauge field on the resonance. In general, a classical limit of the strong field is sensible to be imprinted on the nuclear shape which is described by the IBM symmetries.

\section{Conclusions}
Two classes of results are reported on this paper. The first class refers to the solutions of the scattering problem of two slow neutrons onto a heavy, even-even nucleus at the unitary limit and to the appearance of states in the compound $A+2n$ nucleus that represent conformal symmetry in one dimension - time. The second class refers to the connection with the physical experiment that is realized through fluctuations of cross sections in $A+2n$ compound nuclei. 

The first class of results starts from the solution for the scattering state (s-wave) of the neutron pair at the unitary limit.  The solution of the pair's state in terms of an incident Neummann plus an outgoing Hankel function satisfies the boundary condition at the unitary limit. That boundary condition compensates the unitary pair-collective state interaction in the Schrodinger equation of the pair. The boundary condition depends on the pair-collective state scattering length and is written down explicitly for the channel wavefunction that consists of the product of the pair's state with the IBM target state. 

 The pair-collective state scattering length is tuned to infinity during the resonance of the pair of neutrons with the nearest closed channel IBM state. That resonance is the primary state of the unitary limit with a scaling exponent that is given by the boson number of that closed channel IBM state. Conformal symmetry is represented on the mappings of this primary state to IBM states that represent intermediate states of the $A+2n$ compound nucleus. These mappings apply both to the IBM state of the nearest closed channel and also to higher closed channels that differ by two bosons with respect to the boson number of the primary state. The states of these higher closed channels constitute the excited members of the tower of equally spaced states emerging from the $SO(2,1)$ group - that is isomorphic with the conformal algebra in one dimension - while the IBM state of the nearest closed channel corresponds to the ground member of the tower. 

The second class of results starts from the addition of a second-extra term, apart from the unitary interaction, in the pair-collective state coupling which changes target states and is responsible for the open-closed channel couplings. These couplings produce the width of the intermediate state of the $A+2n$ compound nucleus. The width is experimentally measurable through the fluctuation of the cross section which is represented by the compound-elastic cross section. The fluctuation tunes the pair-collective state scattering length which goes to infinity when the energies of the intermediate states coincide with the energies of the closed channel IBM states.  The representations of conformal symmetry - the tower of equally spaced states - manifest themselves as a regularity pattern of a sequence fluctuations with determined positions and widths. The results are restricted to the fluctuations represented by the compound-elastic reaction. In the limit of a large boson number, the width of each fluctuation is expected to remain constant along the sequence.

 In the tentative application of the above to the process of a two-neutron transfer, the couplings are provided by the creation and subsequent annihilation of one $s$  boson in the target nucleus. This operator is a two-neutron transfer operator for the $A+2n$ compound nucleus and is a primary operator with respect to the $SO(2,1)$ conformal algebra. The intensity of the transfer is now imprinted on the width of the fluctuation and is determined by the product of the boson number, of a spectroscopic factor for the intermediate state and of the pair's momentum.  
 In such a case one determines the widths of each fluctuation in the sequence and is able to define the lower limit in the energy interval for the average values of the cross sections. The sequence of fluctuations is expected to be measured in a series of isotopes with respect to the mass number of the target nucleus in terms of added (subtracted) pairs of neutrons. Such a two-neutron transfer process is proposed for a series of exotic isotopes, close to the $O(6)$ limit, where the incident pairs of neutrons are expected to be emitted back in a relatively short time due to the instability of exotic isotopes.

Some theoretical consequences of the unitary limit in the $A+2n$ compound nucleus are also introduced. The intermediate states at the unitary limit are equivalent with pure IBM states in the continuum in analogy with the bound states in the continuum of single neutron resonances on heavy nuclei. On the other hand, the introduced unitary pair-collective state interaction and scattering length reflect the introduction of a unitary boson-boson interaction and of a boson-boson scattering length in the IBM in analogy with the unitary molecule-molecule interactions in systems of cold atoms. 

On the other hand, the regularity pattern of fluctuations has a corresponding paradigm in Bose-Einstein condensates of systems of cold atoms through the appearance of equally spaced oscillation modes in a condensate \cite{Chevy} that is perturbed via a transverse magnetic field. In a heavy, even-even target nucleus, the transverse excitation corresponds to the two-neutron separation energy. This analogy draws the attention to the manifestation of a Bose-Einstein Condensation limit in the $A+2n$ compound nucleus. To that aspect, the unitary limit corresponds to the BCS-BEC crossover while condensation conditions may be relevant to the evaporation decay in terms of pairs of neutrons in analogy with the evaporative cooling of molecular BECs. The evaporation decay of the $A+2n$ compound nucleus refers to inelastic exit channels which were not included in the present paper.  

To conclude, the unitary limit applies to collective states of heavy even-even nuclei by means of the IBM of nuclear structure and of methods used in systems of cold atoms. It is determined what an experiment should measure - the energies and widths of fluctuations of cross sections in $A+2n$ compound nuclei - for the examination of the unitary limit in collective nuclear states. The regularity pattern of the fluctuations of cross sections is a prediction that differs from their usual random appearance and emerges from the representations of conformal symmetry in $A+2n$ compound nuclei. On its turn, the conformal algebra at the unitary limit produces two main implications. The first is an underlying critical point that manifests itself via the maximization of the fluctuations of the cross section which however retain a finite width and appear in a regularity pattern. The second is the extension of the Lie algebraic framework of the IBM to algebras with infinite number of generators. Apart from the intensive interest of an application of such algebras in compound nuclei, that extension may provide an algebraic method to study the relation of the IBM with a classical limit of the strong interactions.

\acknowledgments{The author is thankful to Pieter Van Isacker for many useful discussions. This research is funded by the European Union's H2020 program, Marie Sklodowska Curie Actions - Individual Fellowships, Grant Agreement No 793900-GENESE 17.} 
\appendix
\section{Pair scattering in O(6)}\label{app}
In general the conjugate momentum $k_{R}$ to the hyperspherical coordinate $R$ is written in terms of $k_{12}=k_{1}-k_{2}$ and $K=(k_{1}+k_{2})/2$ as $k^{2}_{R}=k^{2}_{12}/2+2K^{2}$. A scattering amplitude and cross section in terms of $k_{R}$ reflects pair scattering. The same occurs for $k_{r}$ which is the conjugate momentum with respect to the radial shift $r=R-\rho$. 

Take a wavefunction that satisfies the equation
\begin{equation}\label{A0}
	\left(\frac{\hbar^{2}}{2M}\nabla^{2}_{r}+k_{r}^{2}\right)\Psi_{0}=0.
\end{equation}
A scattering wavefunction $\Psi_{0}$ is written as $\Psi_{0}=\psi_{i}+\psi_{s}$ with $\psi_{i}$ the incident wave and $\psi_{s}$ the scattered wave. 
From the many available references in scattering problems, this analysis follows \cite{Morse}. The scattering wavefunction $\Psi_{0}(r)$ shows the relative motion of the incident pair of neutrons with respect to the target nucleus. The Schrodinger equation of the pair respects the $O(6)$ symmetry due to the kinetic term $T_{r}=-(\hbar^{2}/2M)\nabla^{2}_{r}$ as discussed in section \ref{2}.
\subsection{Hyperspherical Coordinates in pair scattering}\label{app00}
 The Schrodinger equation of the pair in hyperspherical coordinates lives in a six dimensional space. The surface element is $dS=r^{5}d\Omega_{5}$. The angles of $\Omega_{5}$ are the $(\theta_{1},\phi_{1})$,$(\theta_{2},\phi_{2})$ and the angle $\alpha \in [0,\pi/2]$. The latter is defined as the tangent of the ratio $r_{1}/r_{2}$, $\alpha=\tan^{-1} \left( r_{1}/r_{2} \right)$ \cite{Fano, Morse} and
\begin{equation}\label{A03}
\quad d\Omega_{5}=\sin^{2}\alpha \cos^{2} \alpha \sin\theta_{1} \sin\theta_{2}d\theta_{1}d\theta_{2}d\phi_{1} d\phi_{2}d\alpha.
\end{equation}
In this system of coordinates a scattering amplitude at the unit of time at the surface element is $(|f|^{2}/r^{5}) r^{5}d\Omega_{5}$. Therefore the differential cross section is $d\sigma = |f|^{2}d\Omega_{5}$.  The scattering amplitude $f$ is written as $f_{k_{r}}$ for the incident pair of neutrons with relative momentum $k_{r}$ and is a function of the angles $f_{k_{r}}(\alpha,\theta_{1},\theta_{2})$ in the azimuthal symmetry. The surface area is $16\pi^{2} \times \pi/16=\pi^{3}$ as seen from Eq (\ref{A03}). 
At large distances from the reaction the wavefunction is written
\begin{equation}\label{A02}
	\lim_{r\rightarrow \infty}\Psi_{0}(r)=e^{ik^{-}_{r}r}+\frac{f_{k_{r}}}{r^{5/2}}e^{ik^{+}_{r}r},
\end{equation}
with $e^{ik^{-}_{r}r}=e^{-ik_{r}r\cos\alpha}$ and $e^{ik^{+}_{r}r}=e^{ik_{r}r}$. Note that an isotropic wave in six dimensions is accompanied by the factor $r^{-5/2}$. 
The boundary condition for $\psi_{s}$ is
\begin{equation}\label{A00}
\lim_{r \rightarrow \infty}\psi_{s}(r)=\frac{f_{k_{r}}}{r^{5/2}}e^{ik_{r}r}.
\end{equation}\label{A01}
\subsection{Bessel Functions in pair scattering}\label{app0}
This subsection provides the proof that the pair wavefunction is described by a Bessel function in scattering. The hyperspherical equation, without a trap, is written in the $d$ dimensional space with $d=3N$
\begin{equation}\label{A1}
	-\frac{\hbar^{2}}{2M} \left(  \frac{1}{R^{3N-1}} \frac{\partial}{\partial R}R^{3N-1} \frac{\partial}{\partial R} - \frac{ \Lambda}{R^{2}} \right)\Psi_{0}(R)=E \Psi_{0}(R).
\end{equation}
Set $k^{2}_{R}=2M E/\hbar^{2}$ and introduce the transformation $\Psi_{0}(R)=R^{-a}\chi(R)$. The following cases are obtained:
\begin{widetext}
	\begin{equation}\label{A2}
	a=\frac{3N}{2}-1: \qquad- \left( \frac{\partial^{2}}{\partial R ^{2}} + \frac{1}{R} \frac{\partial}{\partial R} - \frac{\Lambda+\left((3N-2)/2\right)^{2}}{R^{2}} \right)\chi(R)=k^{2}_{R} \chi(R),
	\end{equation}
	while
	\begin{equation}\label{A3}
	a=\frac{3N-1}{2}: \qquad- \left( \frac{\partial^{2}}{\partial R ^{2}} - \frac{\Lambda+((3N-1)(3N-3)/4)}{R^{2}} \right)\chi(R)=k^{2}_{R} \chi(R).
	\end{equation}
\end{widetext}
For $N=2$ in equation (\ref{A3}), $a=5/2$ and therefore $\Psi_{0}(R)=R^{-5/2}\chi(R)$ with $\Lambda+15/4$ to be the numerator of the centrifugal term. This wavefunction defines a spherical wave in six dimensions. 

However in scattering one needs the partial wave analysis and the appropriate equation for that reason is Eq (\ref{A2}). For $N=2$, $\Psi_{0}(R)=R^{-2}\chi(R)$ with $\Lambda+4=\lambda(\lambda+4)+4=(\lambda+2)^{2}$ as the numerator of the centrifugal term. A radial shift $r=R-\rho$ is performed that does not affect the angular motion. The wavefunction with respect to $r$ is $\Psi_{0}(r)=r^{-2}\chi(r)$ and Eq (\ref{A2}) reads 
\begin{equation}\label{A2a}
-\left( \frac{\partial^{2}}{\partial r ^{2}} + \frac{1}{r} \frac{\partial}{\partial r} - \frac{\left(\lambda+2\right)^{2}}{r^{2}} \right)\chi(r)=k^{2}_{r} \chi(r).
\end{equation}
 For $z=k_{r}r$, Eq (\ref{A2a}) takes the form of a Bessel equation
\begin{equation}\label{A5}
	\left( \frac{\partial^{2}}{\partial z ^{2}} + \frac{1}{z} \frac{\partial}{\partial z}+1 - \frac{(\lambda+2)^{2}}{z^{2}} \right)N_{\lambda+2}(z)=0,
\end{equation}
with $\chi(r)=N_{\lambda+2}(k_{r}r)$. In general the solutions of Eq (\ref{A5}) are provided by any kind of Bessel functions such as the first kind $J_{\lambda+2}(k_{r}r)$, Neummann $N_{\lambda+2}(k_{r}r)$ or Hankel functions $H^{(1,2)}_{\lambda+2}(k_{r}r)$.  By choosing a normalization factor $C=\sqrt{\pi/2k_{r}}e^{i5\pi/4}$, the solutions of Eq (\ref{A2a}) read
\begin{equation}\label{B1}
\Psi_{0}(r)=C\frac{N_{\lambda+2}(k_{r}r)}{r^{2}}.
\end{equation}		
The asymptotic behavior of Bessel functions are listed in Table \ref{t1}. For the second kind, the behavior of $\Psi_{0}(r)$ away from the reaction center is
\begin{equation}\label{B2}
\lim_{r \rightarrow \infty}\Psi_{0}(r)=\frac{i^{5/2}}{k_{r}r^{5/2}}\sin\left(k_{r}r-\frac{1}{2}\left(\lambda+\frac{5}{2}\right) \pi\right).
\end{equation}	
We are now in the position to perform the partial wave analysis of the incident wavefunction of Eq (\ref{A02}) through a representation label for the angular motion that classifies the states according to the $O(6)$ symmetry. Useful relations in this system of coordinates amenable to $O(6)$ symmetry are provided by Sommerfeld \cite{Sommerfeld}. Keeping the factor $C$, the  expansion of the incident wave $\psi_{i}(r)=exp(-ik_{r}r\cos \alpha)$ in terms of $\lambda$ reads
\begin{equation}\label{B3}
\psi_{i}(r)= 4 \Gamma(2)\sum^{\infty}_{\lambda=0}(\lambda+2)e^{-i \lambda\pi/2}P_{\lambda}(\cos \alpha|4)C\frac{N_{\lambda+2}(k_{r}r)}{r^{2}}.
\end{equation}
$P_{\lambda}(\cos \alpha|4)$ indicates a Gegenbauer Polynomial of index $\lambda$ and $4$ indicates the $4$-sphere spanned by the angles $(\theta_{1},\phi_{1})$,$(\theta_{2},\phi_{2})$. Since $\Gamma(2)=1$ the numerical factor in front of the summation is 4. The relation (\ref{B3}) is the partial wave expansion of the ingoing wave of the relative motion between the pair of neutrons and the target nucleus. Another type of expansion is given by Morse and Feshbach \cite{Morse} which decomposes the ingoing wave in terms of the relative angular momentum of each neutron i.e $l_{1},l_{2}$ with respect to the core.

 The asymptotic behavior of the incident wave is therefore obtained from (\ref{B3}) and the asymptotic behavior of $N_{\lambda+2}(k_{r}r)$ from Table \ref{t1}. This is
\begin{equation}\label{B4}
\begin{split}
& \lim_{r \rightarrow \infty}\psi_{i}(r)= 4\sum^{\infty}_{\lambda=0}(\lambda+2)e^{-i \lambda\pi/2}P_{\lambda}(\cos \alpha|4) \times \\ 
& \frac{i^{5/2}}{k_{r}r^{5/2}}\sin\left(k_{r}r-\frac{1}{2}\left(\lambda+\frac{5}{2}\right) \pi \right).
\end{split}
\end{equation}
Similarly, the scattering wavefunction $\Psi_{0}(r)$ is expanded in partial waves in the form
\begin{equation}\label{B5}
\Psi_{0}(r)=4\sum^{\infty}_{\lambda=0}(\lambda+2)e^{-i \lambda\pi/2}P_{\lambda}(\cos \alpha|4)\frac{\chi_{\lambda+2}(r)}{k_{r}r^{5/2}}.
\end{equation}	
The functional dependence of each partial wave $\chi_{\lambda+2}(r)$ in $\Psi_{0}(r)$ should be asymptotically the same with the asymptotic limit of each partial wave in $\psi_{i}(r)$ up to the phase shifts $\delta_{\lambda}$. This means that the partial wave $\chi_{\lambda+2}(r)$ gives the asymptotic behavior 
\begin{equation}\label{B6}
\lim_{r \rightarrow \infty}\chi_{\lambda+2}(r)=i^{5/2}e^{i\delta_{\lambda}}\sin\left(k_{r}r-\frac{1}{2}\left(\lambda+\frac{5}{2}\right) \pi +\delta_{\lambda} \right).
\end{equation}
The scattered wave $\psi_{s}=\Psi_{0}-\psi_{i}$ is
\begin{equation}\label{B7}
\begin{split}
&\psi_{s}(r)=4\sum^{\infty}_{\lambda=0}(\lambda+2)e^{-i \lambda\pi/2}P_{\lambda}(\cos \alpha|4) \times \\ 
& i^{5/2}\left( \frac{\chi_{\lambda+2}(r)}{k_{r} r^{5/2}} -C\frac{N_{\lambda+2}(k_{r}r)}{r^{2}} \right).
\end{split}
\end{equation}
By performing the subtraction for $r \rightarrow \infty$, the asymptotic behavior of $\psi_{s}$ reads
\begin{equation}\label{B8}
\begin{split}
& \lim_{r \rightarrow \infty}\psi_{s}(r)=4\sum^{\infty}_{\lambda=0}(\lambda+2)e^{-i \lambda\pi/2}P_{\lambda}(\cos \alpha|4) \times \\ & (-1)^{\lambda} (e^{2i\delta_{\lambda}}-1)\frac{e^{ik_{r}r}}{2ik_{r}r^{5/2}}.
\end{split}
\end{equation}
 The radial behavior of the scattered wave is that of the asymptotic behavior of the Hankel function $(C/r^{2})H^{(1)}_{\lambda+2}(k_{r}r)$. 
 
 The wavefunction $\Psi_{0}(r)$ of the problem at hand should be normalized from the $k_{r}$ scale to the energy scale. The normalization of a Bessel function with respect to $k_{r}$ is independent of the dimensionality of that Bessel function with respect to $r$ \cite{Sommerfeld}. Therefore, the Bessel function in six dimensions $\Psi_{0}(r)$ is normalized to energy units by the factor $\sqrt{(2/\pi) k^{2}_{r} dk_{r}/dE}$ \cite{Feshbach}. The dependence of the energy on the wave number is $E(k_{r})=\hbar^{2} k^{2}_{r}/2M$ with $dk_{r}/dE= M/(\hbar^{2} k_{r})$. 
 
 To recapitulate, the $s$-wave of the wavefunction $\Psi_{0}(r)=\psi_{i}+\psi_{s}$ is written in terms of Bessel functions as
 \begin{equation}\label{B81}
 \Psi_{0}(r)=8C \left( \frac{ N_{0+2}(k_{r}r)}{r^{2}}+\frac{(e^{2i\delta_{0}}-1)}{2ik_{r}}\frac{H^{(1)}_{0+2}(k_{r}r)}{r^{2}} \right).
 \end{equation}
 In the asymptotic limit the last expression is
 \begin{equation} \label{B82}
 \lim_{r\rightarrow \infty}\Psi_{0}(r)= e^{i\delta_{0}} u_{0}^{+}(r), \quad u_{0}^{+}(r)=8i^{5/2}\frac{\sin\left(k_{r}r-\frac{5\pi}{4}+\delta_{0}\right)}{k_{r}r^{5/2}}.
 \end{equation}
\subsection{Scattering Amplitude for the pair}\label{app1}
By the  usual definition $S_{\lambda}=e^{2i\delta_{\lambda}}$. The scattering amplitude of the pair is
\begin{equation}\label{B9}
f_{k_{r}} (\alpha|4)=\frac{1}{2ik_{r}}\sum^{\infty}_{\lambda=0}4(\lambda+2)(-1)^{\lambda}(S_{\lambda}-1)P_{\lambda}(\cos \alpha|4).
\end{equation}
Keeping only the term with $\lambda=0$ in the summation, the scattering amplitude of the relative $s$-wave is
\begin{equation}\label{B10}
f_{k_{r}}=\frac{8 e^{i\delta_{0}} \sin\delta_{0}}{k_{r}}=\frac{8}{k_{r}\cot\delta_{0}-ik_{r}}=\frac{8}{\frac{1}{a_{r}(k_{r})}-ik_{r}}.
\end{equation}
 The differential cross section is
\begin{equation}\label{B11}
\frac{d\sigma}{d\Omega}=\abs{f_{k_{r}}}^{2}.
\end{equation}
There is therefore an upper bound on $f_{k_{r}}$
\begin{equation}
|f_{k_{r}}| \leq \frac{8}{k_{r}}.
\end{equation}
The unitary limit is the case of the maximum scattering amplitude namely $|f^{unitary}_{k_{r}}|=\frac{8}{k_{r}}$. The cross section is
\begin{equation}\label{B12}
\begin{split}
& \sigma=(4\pi)^{3}\frac{\sin^{2}\delta_{0}}{k^{2}_{r}}=\frac{(4\pi)^{3}}{k^{2}_{r}} \left(\frac{1}{1+\cot^{2}\delta_{0}} \right)= \\ & \frac{(4\pi)^{3}}{k^{2}_{r}+1/a^{2}_{r}(k_{r})}.
\end{split}
\end{equation} 
\subsection{Cross Sections}\label{appcs}
The generalized scattering length $a_{r}(k_{r})$ is energy dependent with $k^{2}_{r}=2M E/\hbar^{2}$. If a resonance occurs at energy $E_{m}$, then $1/a_{r}(E_{m})=0$. In the neighborhood of $E_{m}$
\begin{equation}\label{B14}
\frac{1}{a_{r}(E)}=0+(E-E_{m})\frac{d}{dE} \left(\frac{1}{a_{r}} \right){\bigg\rvert}_{E_{m}}+ \cdots.
\end{equation}
 The width of the resonance of the incident pair with the collective state is related with the scattering length by the relation 
\begin{equation}\label{B15}
\frac{d}{dE} \left( \frac{1}{a_{r}} \right){\bigg\rvert}_{E_{m}}=\frac{2k_{r}}{\Gamma_{m}},
\end{equation}
where $2$ is put for reasons of convention with the final Breit-Wigner form. 

This is the resonance for the elastic process. The cross section of the elastic process $\sigma_{el}$ consists of the shape-elastic cross section $\sigma_{se}$ plus the compound elastic $\sigma_{ce}$ which receives contributions from closed channels and finally emits back the pair in the entrance channel. In order to present the form of $\sigma_{ce}$ some remarks for inelastic processes are added in order to clarify the relation of closed channels processes with the generalized scattering length. 

Inelastic scattering creates the reaction cross section $\sigma_{r}$ and occurs whenever the scattering wavefunction receives contributions from the closed channels. In such a case,  the total width of the process is $\Gamma_{m}=\Gamma_{s}+\Gamma_{r}$ and receives contributions from the decay amplitude back to the open channel $\Gamma_{s}$ and from that towards the closed channel $\Gamma_{r}$. The wavevector $k_{r}$ is analytically continued to the complex plane. Namely, the energy of the resonance is $E_{m}=\epsilon_{s}+i\Gamma_{r}$. Then the generalised scattering length $a_{r}(k_{r})$ becomes complex too \cite{Bethe}. That is
\begin{equation}\label{B16}
Re\left[\frac{d}{dE}\left( \frac{1}{a_{r}}\right){\bigg\rvert}_{E_{m}} \right]=\frac{2k_{r}}{\Gamma_{s}};  Im\left[\frac{d}{dE}\left( \frac{1}{a_{r}}\right){\bigg\rvert}_{E_{m}} \right]=k_{r} \alpha,
\end{equation}
with $\alpha$ a constant. From these relations, one obtains
\begin{equation}\label{B17}
\begin{split}
& Re\left[ \frac{1}{a_{r}(E)} \right]=\frac{2k_{r}}{\Gamma_{s}}\left[E-\left(\epsilon_{s}+\frac{\alpha \Gamma_{r} \Gamma_{s}}{4}\right) \right], \\
& Im\left[\frac{1}{a_{r}(E)} \right]=k_{r} \left[\frac{\Gamma_{r}}{\Gamma_{s}}+\alpha(E-\epsilon_{s}) \right].
\end{split}
\end{equation}
 At, or very close to the resonance energy $E \sim \epsilon_{s}$, the imaginary part of the scattering length is $Im\left[1/a_{r}(E) \right]=k_{r}\Gamma_{r}/\Gamma_{s}$.

  The total cross section of the process is $\sigma_{tot}=\sigma_{el}+\sigma_{r}$  with their definitions for the pair to be
\begin{equation}\label{B171}
\sigma_{el}=\frac{(4\pi)^{3}}{k_{r}^{2}}|1-S_{0}|^{2}, \quad \sigma_{r}=\frac{(4\pi)^{3}}{k_{r}^{2}}(1-|S_{0}|^{2}).
\end{equation}
In terms of the scattering length they are written as
\begin{equation}\label{B18}
\sigma_{el}=\frac{(4\pi)^{3}}{\abs{ik_{r}+1/a_{r}}^{2}}, \quad \sigma_{r}=\frac{(4\pi)^{3} Im(1/a_{r})}{k^{2}_{r}\abs{ik_{r}+1/a_{r}}^{2}}.
\end{equation}
For the compound elastic reaction cross section $\sigma_{ce}$, one takes the averages of these values and computes the fluctuations 
\begin{equation}\label{B171a}
	\sigma_{ce}=\frac{(4\pi)^{3}}{k_{r}^{2}}\left(\langle|S_{0}|^{2} \rangle -|\langle S_{0} \rangle|^{2} \right).
\end{equation}
For the unitary limit associated with the compound-elastic reaction, the intermediate state has a width $\Gamma_{m} \sim \Gamma_{s}$.

The cross section of the fluctuations around the average value of the pair's cross section are of interest with respect to the measurement of the tower state. This cross is obtained by the difference between the direct or shape-elastic and the total cross section i.e is the compound-elastic reaction. The Breit-Wigner forms of the cross sections are 
\begin{equation}\label{B181}
\begin{split}
& \sigma_{el}=\frac{(4\pi)^{3}}{k^{2}_{r}}\abs{(e^{2i\delta_{0}}-1)+\frac{i\Gamma_{m}}{(E-E_{m})+i(\Gamma_{m})/2}}^{2}, \\ & \sigma_{r}=\frac{(4\pi)^{3}}{k^{2}_{r}} \frac{\Gamma_{s}\Gamma_{r}}{(E-E_{m})^{2}+(\Gamma_{s}+\Gamma_{r})^{2}/4}, \\
& \sigma_{ce}=\frac{(4\pi)^{3}}{k^{2}_{r}} \frac{\Gamma^{2}_{m}}{(E-E_{m})^{2}+(\Gamma_{m})^{2}/4}.
\end{split}
\end{equation}
\section{Boundary condition}\label{appBC}
The solutions to equation (\ref{A2a}) are provided by any kind of Bessel functions or a linear combination of them provided they satisfy the boundary conditions near and far away from the reaction center. Appendix \ref{app0} determined the boundary condition for $r \rightarrow \infty$ and the general form of the phase behavior of the solution at that limit. Now, the boundary conditions that the radial wavefunction is subjected to near the reaction center are the following: (i) When $R \rightarrow \rho$ or $r \rightarrow 0$, the action of the laplacian $\nabla^{2}_{r}$ to the wavefunction $\Psi_{0}(r)$ should reproduce the delta function $\delta(r)$. For the six dimensional space, $\nabla^{2}_{r} (1/r^{4})=-4\pi^{3}\delta(r)$ and the relevant boundary condition is $\lim_{r \rightarrow 0}u_{0}(r)=A/r^{4}+B$. The behavior of $\sim 1/r^{4}$ is not reproduced by the $s$-wave of the Bessel of the first kind $(C/r^{2})J_{0+2}(k_{r}r)$ which goes like $ \sim r^{2}/r^{2}$ for small arguments as seen in Table I. It is satisfied by the Neummann function $(C/r^{2})N_{0+2}(k_{r}r)$ and the Hankel function as well. (ii) A finite range of the pair-collective state interaction can be approximated by taking the collective state as a hard sphere of radius $\rho+a_{r}$ or the pair as a hard sphere of radius $a_{r}$. The hard sphere approximation \cite{Huang} enhances the solution with the condition $u_{0}(a_{r})=0$. This condition characterizes the solution when the range of the interaction is roughly equal with the scattering length. The two conditions are summarized in the form
\begin{equation}\label{e2911}
\lim_{r \rightarrow 0}u_{0}(r)=\frac{A}{r^{4}}+B, \quad A+a^{4}_{r}B=0.
\end{equation}
A simple choice for $u_{0}(r)$ is the Neummann function which for small arguments gives $\lim_{k_{r}r \rightarrow 0} (C/r^{2})N_{0+2}(k_{r}r)=-4C/(\pi k^{2}_{r}r^{4})$. Thus $B=4C/(\pi k^{2}_{r}a^{4}_{r})$ and the solution is
\begin{equation}\label{e2912}
u_{0}(r)=C\left(\frac{N_{0+2}(k_{r}r)}{r^{2}}+\frac{4C}{\pi k_{r}^{2}a^{4}_{r}}\right).
\end{equation}
It is understood that $u_{0}(r)$ is the $r$ part of the open channel consisting of the pair of wavefunctions $\Psi_{0}(r)$ and $\Phi_{0}(\rho)$. The latter may be scaled approriately in order to scale out the factor $4C/(k^{2}_{r}\pi)$. That is the boundary condition at low $r$ is
\begin{equation}\label{B143}
\lim_{r \rightarrow 0}\Psi_{0}(r)\Phi_{0}(\rho)=\Phi_{0}(\rho)\left(-\frac{1}{r^{4}}+\frac{1}{a^{4}_{r}} \right).
\end{equation}
 Now, if the range of the interaction $r^{*}$ is smaller than $a_{r}$ the hard sphere approximation is removed along with the condition (ii). The region of interest is $r^{*}<r<k^{-1}_{r}$. For that region the solutions are normalized to the wavefunction
\begin{equation}\label{e2913}
u_{0}^{+}(r)=8i^{5/2}\frac{\sin\left(k_{r}r-\frac{5\pi}{4}+\delta_{0}\right)}{k_{r}r^{5/2}}.
\end{equation} 
In this case the boundary condition is examined for the wavefunction $e^{i\delta_{0}}u^{+}_{0}(r)=\lim_{r \rightarrow \infty}\Psi_{0}(r)$. The wavefunction $\Psi_{0}(r)$ is written in terms of the incident Neummann plus the scattered Hankel function
 \begin{equation}\label{B81a}
\Psi_{0}(r)=8C \left( \frac{ N_{0+2}(k_{r}r)}{r^{2}}+\frac{(e^{2i\delta_{0}}-1)}{2ik_{r}}\frac{H^{(1)}_{0+2}(k_{r}r)}{r^{2}} \right).
\end{equation}
The limit of the last expression for low $k_{r}r$ is obtained from Table I with $\delta_{0}\sim -k_{r}a_{r}$ and therefore
 \begin{equation}\label{B81b}
\lim_{r \rightarrow 0} \Psi_{0}(r)=\frac{4C}{\pi k^{2}_{r}}\frac{(-1+ia_{r})}{r^{4}}.
\end{equation}
 With the appropriate scaling of $\Phi_{0}(r)$
 \begin{equation}\label{B81c}
\lim_{r \rightarrow 0}\Psi_{0}(\rho,r)=\Phi_{0}(\rho)\frac{(-1+ia_{r})}{r^{4}}.
\end{equation}
\section{Solutions to coupled channels}\label{app11}
 \begin{table}
	\caption{Asymptotic behavior of Bessel functions from \cite{Abramowicz}.}\label{t1}
	\begin{tabular}{c | c | c}
		\hline \hline 
		Bessel & $z \rightarrow 0$ & $z \rightarrow \infty$ \\ \hline
		$J_{\lambda+2}(z)$ & $\frac{1}{(\lambda+2)!}\left(\frac{z}{2} \right)^{\lambda+2}$ & $\sqrt{\frac{2}{\pi z}}\cos\left(z-\frac{1}{2}\pi\left(\lambda+\frac{5}{2}\right)\right)$ \\  
		$N_{\lambda+2}(z)$ &  $-\frac{(\lambda+1)!}{\pi}\left(\frac{2}{z} \right)^{\lambda+2}$ & $\sqrt{\frac{2}{\pi z}}\sin\left(z-\frac{1}{2}\pi\left(\lambda+\frac{5}{2}\right)\right)$  \\
		$H^{(1)}_{\lambda+2}(z)$ &   $i N_{\lambda+2}(z)$ & $\sqrt{\frac{2}{\pi z}}e^{i\left(z-\frac{1}{2}\pi\left(\lambda+\frac{5}{2}\right)\right)}$                       \\
		\hline \hline
	\end{tabular}
\end{table}

For the purposes of the solutions to the coupled channel equations, 
\begin{equation}
\begin{split}\label{e261}
(E-H_{PP})P|\Psi \rangle=H_{PQ} Q |\Psi \rangle,\\
(E-H_{QQ})Q|\Psi \rangle=H_{QP} P|\Psi \rangle,
\end{split}
\end{equation}
the asymptotic limit of the solution $\Psi^{+}_{0}(r)$ is expressed in the form of two isotropical waves in six dimensions
\begin{equation}\label{e2914}
\Psi^{+}_{0}(r)=8\frac{e^{-i(k_{r}r-5\pi/4)}}{r^{5/2}}-S_{0}8\frac{e^{ik_{r}r}}{r^{5/2}},
\end{equation}
with $\Psi^{+}_{0}(r)=-2i k_{r}e^{i\delta_{0}}u^{+}_{0}(r)$. On the other hand if the scattering wavefunction named as $\Psi^{k+}_{0}$ corresponds to an incident "plane" wave plus a scattered wave in the form of a Neummann plus a Hankel function then $\Psi^{k+}_{0}(r)=e^{i\delta_{0}}u^{+}_{0}(r)$. The wavefunction is normalized to the energy scale by the factor
\begin{equation}\label{e2915}
\Psi^{k+}_{0}(r)=\sqrt{\frac{2}{\pi} k^{2}_{r} \frac{dk_{r}}{dE}}e^{i\delta_{0}}u^{+}_{0}(r).
\end{equation}
The solution to the first equation of (\ref{e261}) is given by the expression
\begin{equation}\label{e30}
\langle r|P|\Psi\rangle= \Psi^{+}_{0}(r)|\Phi_{0}(\rho) \rangle+\frac{\langle r|H_{PQ}Q |\Psi \rangle}{E-H_{PP}+i\eta}.
\end{equation}
Call the state of a specific reaction channel as $|c_{n} \rangle \equiv |\Psi_{n} \rangle |\Phi_{n}(\rho) \rangle$. The notation of \cite{Timmermans} for the propagator of the outgoing waves $g^{+}_{p}(E)=(E-H_{PP}+i\eta)^{-1}$ is used below. The amplitude of the open channel state is $\langle r|c^{+}_{0}\rangle = \Psi^{+}_{0}(r) |\Phi_{0}(\rho) \rangle$. The amplitude $\langle r|c^{k+}_{0}\rangle = \Psi^{k+}_{0}(r) |\Phi_{0}(\rho) \rangle$ serves for the expansion of the propagator $g^{+}_{p}(E)$ over the plane waves $\Psi^{k+}_{0}(r)$ in the sense of \cite{Timmermans}.

 By substituting $P|\Psi \rangle $ of Eq (\ref{e30}) into the second equation of (\ref{e261}) one obtains the states of the closed channels
\begin{equation}\label{e31}
Q |\Psi \rangle= \frac{H_{QP}|c^{+}_{0}\rangle}{E-H_{QQ}-H_{QP}g^{+}_{p}(E)H_{PQ}}.
\end{equation}
Putting this back to Eq (\ref{e30}) one takes the solution for the open channel state
\begin{equation}\label{e32}
\begin{split}
& \Psi_{0}(r)|\Phi_{0}(\rho) \rangle= \Psi^{+}_{0}(r)|\Phi_{0}(\rho) \rangle+ \\ &\frac{\langle r | g^{+}_{p}(E)H_{PQ}  \hat{I} H_{QP} |c_{0}^{+} \rangle}{E-H_{QQ}-H_{QP}g^{+}_{p}(E)H_{PQ}}.
\end{split}
\end{equation}
In the second term of the right hand part of the last expression one defines the identity operator $\hat{I}$ of the closed channels space which is spanned by the intermediate states. For the IBM target states, intermediate states are the eigenstates of the operator $H_{QQ}$ determined from the homogeneous part of the second Eq (\ref{e261})
\begin{equation}\label{e33}
(\epsilon_{n}-H_{QQ})\Psi_{n}(r)|\Phi_{n}(\rho) \rangle=0,
\end{equation}
with $n>0$. The identity operator for the closed channels space is $\hat{I}=\sum_{n>0}|c_{n} \rangle \langle c_{n}|$. The diagonal matrix $H_{QQ}$ contains $n\times n$ entries with the $n$-th entry of the main diagonal to give the equation
\begin{equation}\label{e331}
(T_{r}+E_{n})\Psi_{n}(r)|\Phi_{n}(\rho)\rangle=\epsilon_{n}\Psi_{n}(r)|\Phi_{n}(\rho)\rangle.
\end{equation}
 Now, the second term of the right hand part of Eq (\ref{e32}) is evaluated by taking the mean values over the intermediate states. The following expressions are obtained
\begin{equation}\label{e34}
\begin{split}
& \frac{ \hat{I} }{E-H_{QQ}-H_{QP}g^{+}_{p}(E)H_{PQ}} = \sum_{n>0}\frac{ |c_{n} \rangle \langle c_{n}| }{E-E_{n'}+i\frac{\Gamma_{n}}{2}}, \\
& E_{n'}=Re\langle c_{n}|H_{QQ}+H_{QP}g^{+}_{p}(E)H_{PQ}|c_{n} \rangle \equiv \epsilon_{n}+\Delta_{n}, \\
& \frac{\Gamma_{n}}{2}=-Im\langle c_{n}|H_{QP}g^{+}_{p}(E)H_{PQ}|c_{n} \rangle.
\end{split}
\end{equation} 
From the sum over $n$ one chooses that intermediate state with energy closest to the total energy of the open channel, call it $m$. This is the intermediate state of the first closed channel. In order to calculate the width $\Gamma_{m}/2$ one expands the propagator $g^{+}_{p}(E)$ in the continuum as an integral over open channel states with plane waves $\langle r|c^{k+}_{0}\rangle=\Psi^{k+}_{0}(r)$. That is \cite{Feshbach}
\begin{equation}
\Gamma_{m}=2 \pi| \langle c_{m}|H(\rho,R)|c^{k+}_{0}\rangle|^{2}.
\end{equation}
The wavefunction $\Psi^{k+}_{0}(r)$ is normalized to the energy scale by the factor $\sqrt{(2/\pi) k^{2}_{r}dk_{r}/dE}$. This wavefunction absorbs the unitary term of the coupling $H(\rho,R)=g_{r}\delta(r)+F$ by the boundary condition. The second term $F=s^{\dagger}+s$ determines the  magnitude of the open-closed coupling as $H_{m0}=\sqrt{N_{b}+1}$. Therefore
\begin{equation}\label{e340}
\Gamma_{m}=4{\bigg\rvert}\int dr \Psi_{m}(r) H_{m0}u^{+}_{0}(r){\bigg\rvert}^{2} \times \left( k^{2}_{r}  \frac{dk_{r}}{dE} \right).
\end{equation}
Since $dk_{r}/dE= M/(\hbar^{2} k_{r})$, the 
width is simply
 \begin{equation}\label{e3400}
 \Gamma_{m}=b^{2} \left(\frac{4 M} {\hbar^{2}}\right) k_{r} ,
 \end{equation}
with the coupling $b \equiv \sqrt{N_{b}+1} \int dr \Psi_{m}(r)u^{+}_{0}(r)$ to be determined by the boson number times the overlap of the scattering wavefunction with the intermediate state.

One turns now to the calculation of the nominator of Eq (\ref{e32}). The propagator here is a Green function $g^{+}_{p}(E;r,r')$ from the point near the reaction center $r'$ to the point $r$ away from that. It is expanded as
\begin{equation}\label{e301}
g^{+}_{p}(E;r,r')=\frac{\langle r|c^{k+}_{0}\rangle\langle c^{k+}_{0}|r' \rangle}{E-H_{PP}+i\eta}.
\end{equation}
This Green function is given \cite{Morse} as the product of the Neummann function times the Hankel function. Here the asymptotic approximation of \cite{Timmermans} is adopted. That is the propagator is expressed as the asymptotic expression of the Neummann function at $r'$,  $e^{i\delta_{0}}u^{+}_{0}(r')$ times the asymptotic expression of the Hankel function at $r$, $8e^{ik_{r}r}/r^{5/2}$ times the available phase space factor $4M/\hbar^{2}$. With these asymptotic forms, the propagator is
 \begin{equation}\label{e302}
 \lim_{r \rightarrow \infty}g^{+}_{p}(E;r,r') =-\frac{4 M} {\hbar^{2}}8\frac{e^{ik_{r}r}}{r^{5/2}}e^{i\delta_{0}}u^{+}_{0}(r')|\Phi_{0}(\rho) \rangle \langle \Phi_{0}(\rho)|.
 \end{equation}
 Having obtained the expression for the propagator, the nominator of Eq (\ref{e32}) amounts to the evaluation of the expression $\langle r|g^{+}_{p}(E) H_{PQ}  |c_{m} \rangle\langle c_{m}| H_{QP} |c_{0}^{+} \rangle$ far away from the reaction center. The matrix element $\langle c_{m}| H_{QP} |c_{0}^{+} \rangle$  is written as $-2ik_{r}e^{i\delta_{0}}\sqrt{N_{b}+1} \int dr \Psi^{*}_{m}(r)u^{+}_{0}(r)$. Therefore $\langle c_{m}| H_{QP} |c_{0}^{+} \rangle=-2ik_{r}e^{i\delta_{0}}b$
and
\begin{equation}\label{e304}
\begin{split}
\lim_{r\rightarrow \infty} g^{+}_{p}(E; r,r') H_{PQ}  |c_{m} \rangle=-\frac{4 M} {\hbar^{2}}8\frac{e^{ik_{r}r}}{r^{5/2}}e^{i\delta_{0}}b|\Phi_{0}(\rho) \rangle.
\end{split}
\end{equation}
The result for the nominator of the second term of Eq (\ref{e32}) is
\begin{equation}\label{e305}
\begin{split}
& \lim_{r\rightarrow \infty} \langle r|g^{+}_{p}(E) H_{PQ}  |c_{m} \rangle\langle c_{m}| H_{QP} |c_{0}^{+} \rangle= \\ & ie^{2i\delta_{0}}b^{2}\frac{4 M} {\hbar^{2}}k_{r}8\frac{e^{ik_{r}r}}{r^{5/2}}|\Phi_{0}(\rho) \rangle=ie^{2i\delta_{0}}\Gamma_{m}8\frac{e^{ik_{r}r}}{r^{5/2}}|\Phi_{0}(\rho) \rangle.
\end{split}
\end{equation}
Therefore, the open channel wavefunction with the contribution of the intermediate state $m$ reads
\begin{equation}\label{e321}
\begin{split}
& \Psi_{0}(r)|\Phi_{0}(\rho) \rangle= \Psi^{+}_{0}(r)|\Phi_{0}(\rho) \rangle+ \\ & ie^{2i\delta_{0}}\frac{ \Gamma_{m}}{E-E_{m'}+i\frac{\Gamma_{m}}{2}}8\frac{e^{ik_{r}r}}{r^{5/2}}|\Phi_{0}(\rho) \rangle.
\end{split}
\end{equation}
By substituting $\Psi^{+}_{0}(r)$ of Eq (\ref{e2914}) the amplitude of the channel state $\langle r|c_{0} \rangle$ reads
 \begin{equation}\label{e322}
 \begin{split}
 & \Psi_{0}(r)=8i^{5/2}\frac{e^{-ik_{r}r}}{r^{5/2}}-  \left(1-\frac{i \Gamma_{m}}{E-E_{m'}+i\frac{\Gamma_{m}}{2}}\right)e^{2i\delta_{0}}8\frac{e^{ik_{r}r}}{r^{5/2}}.
 \end{split}
 \end{equation}
 \subsection{Effective scattering length}
  In general, for low $k_{r}$, $S_{0}=e^{2i\delta_{0}}=e^{-2ik_{r}a_{r}}$. However, in view of (\ref{e322}) the presence of the intermediate state $m$ changes the quantity $S_{0}$ to the form 
 \begin{equation}\label{B19}
 S_{0}=e^{2ik_{r}a_{r}}\left(1-i\frac{\Gamma_{m}}{E-E_{m'}+i\Gamma_{m}/2} \right).
 \end{equation}
 The parenthesis leads to the definition of the effective scattering length $a_{reff}=a_{r}+a'_{r}$ that includes the width $\Gamma_{m}$ of the intermediate state with
 \begin{equation}\label{B20}
 e^{2ik_{r}a'_{r}}=1-i\frac{\Gamma_{m}}{E-E_{m'}+i\Gamma_{m}/2},
 \end{equation}
 and
 \begin{equation}\label{B21}
 a_{reff}=a_{r}+\frac{1}{2k_{r}}\tan^{-1}\left(\frac{\Gamma_{m}(E-E_{m'})}{(E-E_{m'})^{2}+\Gamma_{m}^{2}/4} \right).
 \end{equation}

\end{document}